\pdfoutput=1
\documentclass[11pt,fleqn]{article}
\usepackage[normalem]{ulem}
\usepackage{amsmath, amssymb,amsthm,cite}
\usepackage[dvips]{graphics}
\usepackage[dvipsnames]{xcolor}
\usepackage{graphicx}
\usepackage{tabulary}
\usepackage{float}
\usepackage{authblk}
\usepackage[font=small,labelfont=bf]{caption}
\usepackage{subcaption}
\usepackage{tikz}
\usepackage[hyperfootnotes=false]{hyperref}
\hypersetup{ 
 colorlinks=true, 
 citecolor=ForestGreen,
 linkcolor=NavyBlue,
 urlcolor=Violet}

\usepackage{physics} 

\DeclareGraphicsExtensions{.pdf,.png,.jpg,.eps}
\usepackage{mathtools}
\usepackage{epsfig}
\usepackage{latexsym}
\usepackage{amsmath}
\usepackage{amssymb}
\usepackage{mhsetup}
\usepackage{mathtools}
\usepackage{xfrac}
\numberwithin{equation}{section}
\numberwithin{table}{section}
\numberwithin{figure}{section}
\usepackage{tabulary}
\usepackage{threeparttable}
\usepackage{array}
\usepackage{booktabs}
\usepackage{multirow}
\usepackage{pbox}
\usepackage{lipsum}
\usepackage{enumerate}
\usepackage{caption,setspace}
\title{Nonlinear Incompressible Shear Wave Models in Hyperelasticity and Viscoelasticity Frameworks, with Applications to Love Waves}
\author[1]{Shawn Samuel Carl McAdam\thanks{Electronic mail: ssm259@usask.ca}}
\author[1]{Samuel Opoku Agyemang\thanks{Electronic mail: soa964@usask.ca}}
\author[1]{Alexei Cheviakov\thanks{Corresponding Author. Alternative English spelling: Alexey Shevyakov. \\ Electronic mail: alexei.cheviakov@usask.ca}}
\affil[1]{Department of Mathematics and Statistics, University of Saskatchewan}

\def\beq{\begin{equation}}
\def\eeq{\end{equation}}

\def\const{\hbox{\rm const}}

\def\sign{\mathop{\hbox{\rm sign}}}

\def\div{{\hbox{\rm div}}}

\def\vec#1{{\boldsymbol{\rm #1}}} 
\def\tens#1{{\boldsymbol{\rm #1}}} 

\newcommand\diff[1][]{\,\mathrm{d}}

\newtheorem{theoremp}{Principal Result}

{\theoremstyle{definition}

}

\newcounter{tabnum}

\renewcommand{\div}{\mathrm{div}}

\begin{document}
\maketitle \numberwithin{equation}{section}
\maketitle \numberwithin{remark}{section}
\numberwithin{lemma}{section}
\numberwithin{proposition}{section}
\begin{abstract}
General equations describing shear displacements in incompressible hyperelastic
materials, holding for an arbitrary form of strain energy density function, are
presented and applied to the description of nonlinear Love-type waves
propagating on an interface between materials with different mechanical
properties. The model is valid for a broad class of hyper-viscoelastic
materials. For a cubic Yeoh model, shear wave equations contain cubic and
quintic differential polynomial terms, including viscoelasticity contributions
in terms of dispersion terms that include mixed derivatives $u_{xxt}$ of the
material displacement. Full (2+1)-dimensional numerical simulations of waves
propagating in the bulk of a two-layered solid are undertaken and analyzed with
respect to the source position and mechanical properties of the layers.
Interfacial nonlinear Love waves and free upper surface shear waves are
tracked; it is demonstrated that in the fully nonlinear case, the variable wave
speed of interface and surface waves generally satisfies the linear Love wave
existence condition $c_1 < \abs{v} < c_2$, while tending to the larger material
wave speed $c_1$ or $c_2$ for large times.
\end{abstract}

\section{Introduction}
The study of nonlinear shear waves propagating through an elastic material is
fundamentally important in various areas of science and technology, including
material sciences, seismology, geology, natural resource exploration,
biological material research, and further medical applications. To briefly
review, an elastic material resists external forces and returns to its original
form when such forces are removed (unless the forces are too extreme, causing the
material to yield and deform plastically). Commonly, models of elastic
materials assume Hooke's law, where stress is linearly related to strain. This
assumption holds for most materials undergoing small deformations, but neglects
nonlinear effects in the stress-strain relation. The present work considers
finite deformations of hyperelastic materials that possess a scalar strain
energy density function $W^H$ determining the stress-strain relationship
(Section \ref{sec:hypervisco}).

Materials such as polymers, rubbers, foams, and biological tissues exhibit
essential nonlinear elastic behaviour when they undergo large elastic
deformations. This nonlinear elastic behaviour under load or prescribed
displacement can be modeled through a phenomenological approach
\cite{marckmann2006comparison}, physical description approach
\cite{treloar1975physics}, or use of experimental data \cite{ogden1972large}.
The strain energy density is commonly either postulated from theoretical
considerations or fitted in a certain form from experimental measurements. The
form of the strain energy density is stated in terms of the deformation
invariants such as principal stretches or matrix invariants of Cauchy-Green
deformation tensors. In addition to hyperelastic behaviour, one may include
viscoelastic constitutive model contributions through, for example, a
pseudo-strain energy density function depending on the rate of change of a
deformation tensor \cite{pioletti2000non}.

A major application area of linear elasticity theory is seismology. The
fundamentals of seismic wave theory are built upon the seismic wave equation
modelling a displacement $\vec{u}(x^1,x^2,x^3) = (u^1,u^2,u^3)$ propagating
through a linear, roughly isotropic material, such as Earth layers and
geological formations:
\begin{equation}
	\label{eqn:seismic_wave}
	\rho\pdv[2]{u^i}{t} = (\lambda+\mu)\pdv{}{x^i}\,\mathrm{div}\,\vec{u} + \mu\,\Delta u^i,
\end{equation}
where the Lam\'e parameters $\lambda,\mu > 0$ and the density $\rho > 0$ may be
assumed piecewise constant \cite{Schettino2015}. The seismic wave equation
follows from the equations of motion of a continuum body and Hooke's law for
isotropic materials. Taking $\lambda,\mu,$ and $\rho$ to be piecewise constant
allows one to model some of Earth's anisotropy, typically as a stack of
isotropic layers. The full seismic wave equation takes the Lam\'e parameters
to be piecewise smooth functions of $(x^1,x^2,x^3)$, adding complexity to the
equation. Using equation \eqref{eqn:seismic_wave}, one can determine that the
divergence and curl of $\mathbf{u}$ are described by D'Alembert's equation (the
linear wave equation) with wave speeds $\sqrt{(\lambda+2\mu)/\rho}$ and
$\sqrt{\mu/\rho}$ respectively. Compressional P--waves are solutions to the PDE
for $\mathrm{div}\,\vec{u}$ and transverse S--waves are solutions to the PDE
for $\mathrm{curl}\,\vec{u}$. In particular, P--waves travel faster than
S--waves.

P-- and S--waves are body waves, and their energy decays as $1/r^2$, where $r$
is the distance from the wave front to the focus of the initial displacement
that caused them, such as an earthquake. Surface waves travel along the surface
of the Earth at a slower rate than body waves, but preserve more of their
energy, decaying as $1/r$. As such, surface waves dominate seismograms at large
distances from an earthquake's focus and cause a majority of the damage to
human structures. One type of surface wave is the Love wave.

Love waves are horizontally polarized transverse waves, referred to as shear or
SH-waves. Over large spatial scales (too large for the flat Earth approximation
to apply), Love waves form from the interference of S--waves repeatedly
reflecting off the slowly curving surface of the Earth \cite{Schettino2015}. A
simplified model of Love waves considers two isotropic, homogeneous materials
in the plane $\mathbb{R}^2$, that is, the flat-Earth model. The first material
is represented by a horizontal strip of thickness $L$, and overlays the second
material, occupying the space below a horizontal line, say $z=0$. We refer to
the line $z=0$ as the interface, and the line $z=L$ as the surface. When the
materials possess differing Young's moduli, displacements between them travel
at differing rates $c_1$ and $c_2$ respectively. Love waves are taken to be the
waves that result from the interactions along the interface of the two media.
Section \ref{sec:linearlove} reviews this model in more detail, and includes an
illustration of the domain (Figure \ref{fig:setup}).

In the fully linear case, it is known that Love waves only exist when $c_1 <
c_2$, which we call the Love wave existence condition. The phase velocity $v$
of a Love wave satisfies
\begin{equation}
	\label{eq:c1c2cL}
  c_1 < \abs{v} < c_2.
\end{equation}
Consequently, for shear waves propagating through linear elastic materials, the
discontinuous wave speed determines the range of allowable phase velocities.
If a seismic ray attempts to cross the interface but its angle of incidence is
too large, then the ray is completely reflected (by Snell's law). Such seismic
rays are then trapped in the upper layer in perpetuity, making the upper
layer a wave guide for seismic rays with large angles of incidence. If $c_2 <
c_1$, such a reflection is impossible so every seismic ray eventually decays
into the half space.

After assuming the appropriate conditions to use the flat Earth approximation
(i.e. considering a rectangle with one face on the Earth's surface and tens of
kilometers deep), we may apply the above Love wave model with the continental
crust as the upper layer, the solid uppermost mantle as the half-space, and the
Mohorovi\v ci\'c (Moho) discontinuity as their interface. Below the oceanic
floor, the Moho is consistently about $10$ km deep. Below continental land
mass, the Moho can range between $20$ km and $70$ km in depth, but averages
about $40$ km deep. We may consider the Moho to be of constant depth whenever
the flat Earth approximation applies.


In the present work, we use continuum mechanics to derive an equation modelling
nonlinear SH waves, that is, waves of finite amplitude, propagating through a
hyperelastic, isotropic, incompressible, homogeneous, neo-Hookean material. The
model generally allows for significant freedom through the choice of the
constitutive strain energy density function $W^H$ to approximate nonlinear
stress-strain curves. We specialize the model to a particular strain energy
density similar to those suggested in Murnaghan \cite{murnaghan1951finite} and
Bland \cite{bland1969nonlinear}. The resulting model only requires a good
approximation of the neo-Hookean material's stress-strain curve with a cubic
differential polynomial. Further, the model is a consequence of three necessary
balance laws in a continuum body: conservation of mass and angular momentum,
and the balance of linear momentum. The resulting model is given by the linear wave
equation with cubic and quintic perturbations, and reduces to the linear model
of SH waves for infinitesimal displacements. Further, our model includes the
effects of energy dissipation through the use of a viscosity potential.

It is important to note that in modelling multiple shear wave settings,
including earthquakes in particular, it is essential to consider finite
displacements and resulting nonlinear PDEs. As verified in Bataille and Lund
\cite{bataille1982nonlinear} the displacements and strains in earthquakes are
small: the typical displacements caused by an earthquake ranges from $10^{-10}
\mathrm{m}$ to $10^{-1}\mathrm{m}$ and typical earthquake strain of the ground
ranges from $10^{-6}$ to $10^{-2}$; however, Bataille and Lund also argued that
nonlinear effects should be included when modelling surface waves because such
nonlinearities may balance out the dispersion effects in current linear Love
wave models. Further, the displacements due to seismic waves are large near the
focus of an earthquake. Because the flat Earth approximation is only valid
over small spatial scales (hence, near the focus of the earthquake), our
nonlinear model of SH waves is ideal for the flat Earth model of Love waves.
Other applications include the description of shear waves in biological tissues
and other complex materials (see, e.g., \cite{cheviakov2016one}).

The remainder of this paper is organized as follows. In Section
\ref{sec:linearlove}, we review the Linear theory of Love waves and the
corresponding solutions. In Section \ref{sec:hypervisco}, we review the
equations of motion and constitutive models of fully nonlinear hyperelastic
bodies in the material (Lagrangian) framework. In Section
\ref{Sec:NonlinearHyperelasticLoveWaves}, we derive a general partial
differential equation model describing the propagation of one- and
two-dimensional shear waves in incompressible hyperelastic materials for an
arbitrary strain energy density function. In Section \ref{sec:NonlinearVisco},
we generalize the work in Section \ref{Sec:NonlinearHyperelasticLoveWaves} to
include viscoelastic effects. In particular, we derive the hyper-viscoelastic
equation of motion satisfied by shear waves in the cubic Yeoh model framework.
In Section \ref{sec:numerical}, we numerically analyze the qualitative
behaviour of solutions to the Love wave equation with finite displacements. In
particular, wave fronts are efficiently tracked, and it is shown that, similar
to the linear framework, the nonlinear Love-type wave speeds generally satisfy
the inequality \eqref{eq:c1c2cL}. In Section \ref{sec:1d}, we use symmetry
methods to derive exact scaling-invariant solutions to our model in one spatial
dimension. Though unbounded, with certain modifications, these solutions
provide close agreement to the observed numerical behaviour.

\section{Linear theory of Love waves}
\label{sec:linearlove}

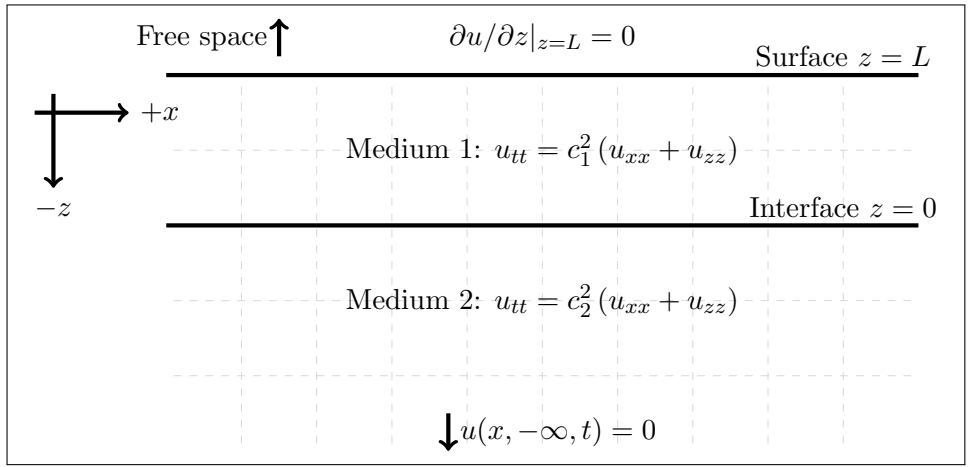
\begin{figure}[htbp]
\centering
\fbox{
\begin{tikzpicture}
\draw[help lines, color=gray!30, dashed] (-4.9,-1.9) grid (4.9,2.9);
\draw[->,ultra thick] (-6.75,2.50)--(-5.5,2.5) node[right]{$+x$};
\draw[->,ultra thick] (-6.50,2.75)--(-6.5,1.5) node[below]{$-z$};
\draw[->,ultra thick] (-1.25,-1.5)--node[right]{$u(x,-\infty,t)=0$}(-1.25,-2);
\draw[->,ultra thick] (-3.5,3.25)--node[left]{Free space}(-3.5,3.75);
\draw[-, ultra thick] (-5,3)--(5,3);
\draw[-, ultra thick] (-5,1)--(5,1);
\node (n1) at (0,2) {Medium $1$: $u_{tt} = c_1^2\left(u_{xx} + u_{zz}\right)$};
\node (n2) at (0,0) {Medium $2$: $u_{tt} = c_2^2\left(u_{xx} + u_{zz}\right)$};
\node (n3) at (4,3.25) {Surface $z=L$};
\node (n3) at (4,1.25) {Interface $z=0$};
\node (n3) at (0,3.5) {$\partial u/\partial z|_{z=L} = 0$};
\end{tikzpicture}
}

\caption{An isotropic layer overlying an isotropic elastic half-space. The $y$
direction is perpendicular to this space spanned by $x,z$. The positive $z$
direction points downwards, $L$ is the depth of the isotropic layer, waves
travelling through the isotropic layer (medium $1$) have speed $c_1^2$, and
waves travelling through the isotropic elastic half-space (medium $2$) have
speed $c_2^2$.}
\label{fig:setup}
\end{figure}

In two-dimensional space, consider an isotropic layer of height $L$ overlying
an isotropic elastic half-space and label their interface with the line $z=0$.
Let $u = u(x,z,t)$ be a displacement perpendicular to this space at the point
$(x,z)$ and time $t$. Suppose the displacement $u$ propagates through the space
according to the following variation of the linear wave equation
\begin{equation}
	\label{eqn:linearlove}
	u_{tt} = c(z)^2\left(u_{xx} + u_{zz}\right), \quad
	c(z) = \begin{cases}
		c_1 & \text{if } 0 \leq z < L, \\
		c_2 & \text{if } z < 0.
	\end{cases}
\end{equation}
where subscripts denote partial differentiation whenever appropriate. We refer
to equation \eqref{eqn:linearlove} with piecewise constant wave speed as the linear Love wave equation. This
equation models materials whose elements have a linear restoring force
resulting from Hooke's law. The wave speed $c^2$ is taken to be a
piecewise-constant function of $z$ because the speed of propagation depends on
various properties of the two media. Love waves will propagate between the interface
and the upper boundary $z=L$ (which we refer to as the surface of this domain).
The surface is stress-free, corresponding to a Neumann boundary condition
$u_z(x,L,t) = 0$. At $z \to -\infty$, $u\to 0$ is assumed. Figure
\ref{fig:setup} illustrates this model.

Commonly, one solves this problem by assuming that for fixed $z$, a Love wave
has the form of a plane wave travelling along the $x$ direction \cite{graff2012wave},
\begin{equation}
	\label{eqn:linearlove_ansatz}
	u(x,z,t) = f(z)e^{ik(x-vt)},
\end{equation}
where $k$ is the wave number, $v$ is the phase velocity of the wave, and $f$ is
some continuously differentiable function of the depth $z$. Here we aim to solve for
$f$ and find an expression for $k$ as a function of $v$. Note that for
\eqref{eqn:linearlove_ansatz} to represent a Love wave, $f$ must decay
exponentially as $z\to-\infty$. Further, observe that the phase velocity is
assumed to not depend on depth $z$. Substituting \eqref{eqn:linearlove_ansatz}
into the piecewise linear wave equation implies
\begin{equation}
	\label{eqn:loveBVP}
	f''(z) = k^2(1-(v/c(z))^2)f(z) =
	\begin{cases}
		k^2(1-(v/c_1)^2)f(z) & \text{if } 0 \leq z < L, \\
		k^2(1-(v/c_2)^2)f(z) & \text{if } z < 0,
	\end{cases}
\end{equation}
and the boundary conditions imply $f'(L) = f(-\infty) = 0$.
For any fixed value of $v$, equation \eqref{eqn:loveBVP} becomes a
Sturm--Liouville boundary value problem (BVP)
\[
	(p(z)f'(z))' + q(z)f(z) = -\lambda r(z)f(z),
\]
where $p=1, q=0, \lambda = k^2$, and $r(z)=1-(v/c(z))^2$. Recall that for an
Sturm--Liouville BVP to be \emph{regular}, the weight function $r(z)$ must be a
positive, continuous function over its domain. The ODE in equation
\eqref{eqn:loveBVP} is \emph{irregular} because $r(z)$ is discontinuous at $0$.

For every $\abs{v}\in(c_1,c_2)$ and every $n\in\mathbb{Z}$, the following is a
countable set of eigenvalue/function pairs $(k_n(v),f_n(z;v))$ for the above
Sturm--Liouville BVP:
\begin{equation}
	\label{eqn:linearlove_soln}
	f_n(z;v) =
	\begin{cases}
		\cos(k_n(v)\Omega_1(v)(z-L))                   & 0 \leq z < L,    \\
		\cos(k_n(v)\Omega_1(v)L)e^{\abs{k_n(v)}\Omega_2(v)z} & z < 0,
	\end{cases}
\end{equation}
where
\[
	\Omega_1(v) = \sqrt{(v/c_1)^2-1}, \quad \Omega_2(v) = \sqrt{1-(v/c_2)^2}.
\]
The derivative $f_n'(z;v)$ must be continuous at $L$, so $\Omega_1(v)$ and
$\Omega_2(v)$ are related by
\begin{equation}
	\label{eqn:condition}
	\Omega_2(v) = \sign(k_n(v))\Omega_1(v)\tan(k_n(v)\Omega_1(v)L).
\end{equation}
Both functions $\Omega_1(v), \Omega_2(v)$ must output nonnegative real numbers
for the solution $f_n$ to satisfy the boundary conditions and be continuously
differentiable at the interface $z=0$. This determines the range of allowable
phase velocities $v$, also called the \emph{Love wave existence condition}
(\ref{eq:c1c2cL}):
\beq\label{eq:LW:ex:cond}
	c_1 < \abs{v} < c_2.
\eeq
The Love wave
existence condition also implies the weight function $r(z) \leq 0$ for
$z\in[0,L)$, which is another reason why the Sturm--Liouville BVP
\eqref{eqn:loveBVP} is not regular.

One can use equation \eqref{eqn:condition} to solve for each eigenvalue $k_n(v)$,
\begin{align*}
	k_n(v) &= \frac{1}{L\Omega_1(v)}\left(\sign(n)\arctan(\Omega_2(v)/\Omega_1(v)) + \pi n\right) \\
			   &= \frac{c_1}{L\sqrt{v^2-c_1^2}}\left(\sign(n)\arctan\left(\frac{c_1}{c_2}\sqrt{\frac{c_2^2-v^2}{v^2-c_1^2}}\right) + \pi n\right) \\
				 &= \sign(n)k_0(v) + \frac{\pi n}{L\Omega_1(v)}.
\end{align*}
Because $k_0(v)$ is nonzero, there are two distinct eigenvalues $\pm k_0(v)$
corresponding to $0$, which we label with $\pm k_0(v)=k_{\pm0}(v)$. We observe
that $k_{-n}(v) = -k_n(v)$ for every $n$ (including $n=0$) so the
eigenfunctions satisfy $f_{-n}(z,v)=f_n(z,v)$. Each eigenfunction then
corresponds to two distinct eigenvalues $\pm k_n(v)$. The negative $n$
eigenfunctions do not yield new linearly independent solutions; however, the
overall solution mode $f_n(z,v)e^{ik_n(v)(x-vt)}$ \emph{does} change. Now that
we have an equation for $k_n$, we see that the inequalities are strict in
\eqref{eq:c1c2cL} because $f_n(z,c_2)=(-1)^n$ for $z<0$ (hence does not decay
as $z\to-\infty$) and $f_n(z,c_1)$ is not continuously differentiable at $z=0$.
Indeed, $f_n(z,c_1)=0$ for $z\leq0$ (because $k_n$ has a pole at $c_1$) and
$\partial_zf_n(z,c_1)|_{z=0^+}=-k_n(c_1)\Omega_1(c_1)\sin(k_n(c_1)\Omega_1(c_1)L)\neq0$.

There is a countably infinite set of wave numbers $k_n(v)$ corresponding to a
given phase velocity $v$. We note that $k_n(v),\Omega_1(v),\Omega_2(v),$ and
$f_n(z;v)$ are all \emph{even} with respect to $v$, so we need only study these
function for positive $v$. Each $k_n(v)$ is a strictly decreasing
function of $v\in(c_1,c_2)$ for $n\geq0$ because $\Omega_1(v)$ strictly
increases, $\Omega_2(v)/\Omega_1(v)$ strictly decreases, and $\arctan$
strictly increases. As such, $k_n(v)$ realizes its minimum at $v=c_2$ and we
have the following lower bound on each $k_n(v)$,
\[
	k_n(v) \geq k_0(c_2) + \frac{\pi n}{L\Omega_1(c_2)} \geq n\frac{\pi c_1}{L\sqrt{c_2^2 - c_1^2}}.
\]
This result implies there are finitely many phase velocities $v$ corresponding
to a given wave number $k^*$. Further, $k_n$ restricted to $[c_1,c_2]$ is
invertible and its inverse has the domain
$[\left. n\pi c_1 \middle/ L\sqrt{c_2^2-c_1^2}\right.,\sign(n)\infty]$.

\begin{figure}[htbp]
\includegraphics[width=\linewidth]{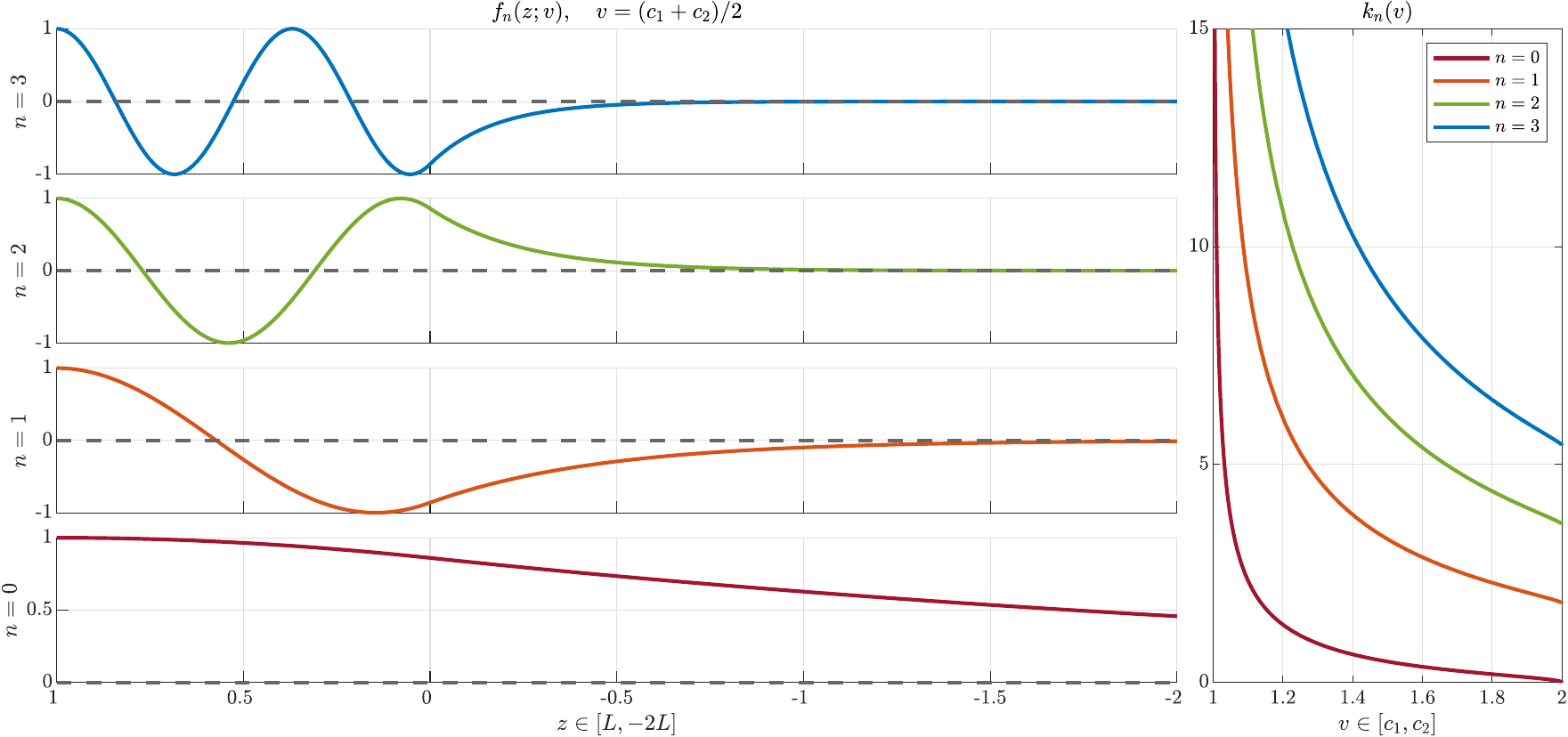}
\footnotesize
\centering

\caption{Left: plots of $f_0(z; v)$, $f_1(z; v)$, $f_2(z; v)$, and
$f_3(z; v)$ for $z\in[0,3L]$ and $v=(c_1+c_2)/2$. Right: plot of
$k_n(v)$ over $v\in[c_1,c_2]$ for $n=0,1,2,3$. Recall that $k_n(v)$ strictly
increases with $n$. Other parameters are $L=1$, $c_1 = 1$, $c_2 = 2$}
\label{fig:plot_kv}
\end{figure}

\begin{figure}[htbp]
\includegraphics[width=\linewidth]{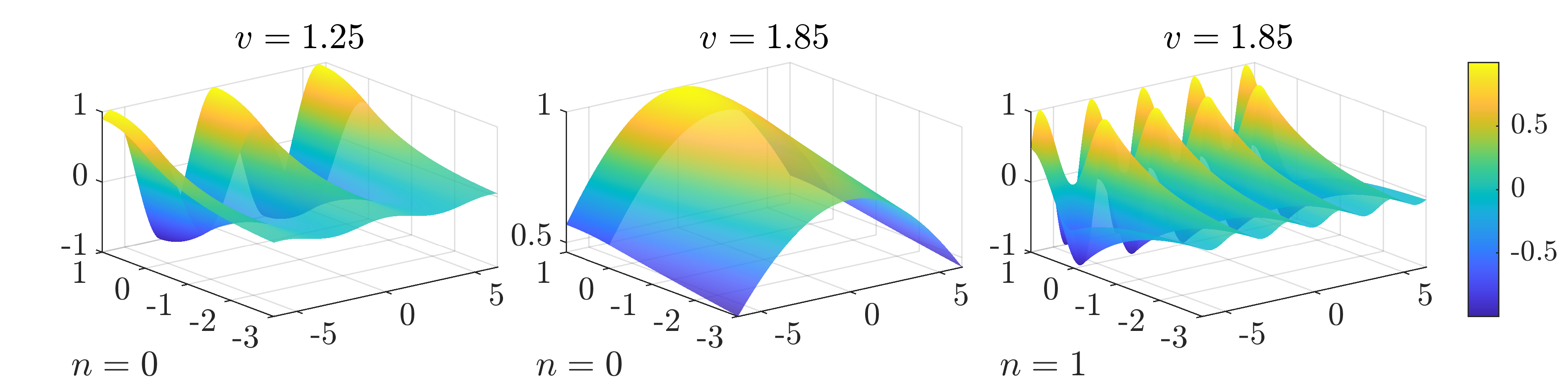}
\footnotesize
\centering
\caption{
The real part of fundamental solutions $f_n(z;v)e^{ik_n(v)(x-vt)}$ to equation
\eqref{eqn:linearlove} for various $v$ and $n$, with $c_1=1$, $c_2 = 2$, and $L = 1$.
}
\label{fig:manufacturedLove_plots}
\end{figure}

Figure \ref{fig:plot_kv} plots $f_n(z; v')$ for $x\in[0,3L]$ with
$v'=(c_1+c_2)/2$, and $k_n(v)$ for $v\in[c_1,c_2]$ both for various values of
$n$. Increasing $n$ by $1$ increases $k_n(v)$ by $\pi/(L\Omega_1(v))$. This
has the effect of increasing the frequency of oscillations of $f_n(z; v)$ for
$z\in[0,L]$, and makes $f_n(z; v)$ decay faster for $z\geq L$. Increasing $n$
also makes $u(x,z,t) = f_n(z;v)e^{ik_n(v)(x-vt)}$ oscillate more along the $x$
direction. This is illustrated in Figure \ref{fig:manufacturedLove_plots} which
contains plots of fundamental solutions to equation \eqref{eqn:linearlove} for
various $n$ and $v$.

A solution of the form $f_n(z; v)e^{ik_n(v)(x-vt)}$ exists for each
$\abs{v}\in(c_1,c_2)$ and each $n\in\mathbb{Z}$. Let
$\mathcal{D}=(-c_2,-c_1)\cup(c_1,c_2)$, then the most general plane wave
solution to the linear Love wave equation \eqref{eqn:linearlove} is of the form
\begin{equation}
	\label{eqn:total_linear_solution}
	u(x,z,t) = \sum_{n=-\infty}^\infty \int_{\mathcal{D}} \alpha_n(v)f_n(z; v)e^{ik_n(v)(x-vt)} \diff{v}
\end{equation}
where we recall that the solution corresponding to $n=0$ is counted twice in
the first line because there are two distinct $0^{\text{th}}$ modes with
eigenvalues $\pm k_0(v)$.

We determine the coefficients $\alpha_n(v)$ from sufficiently smooth initial
conditions $u(x,z,0) = F(x,z)$ and $u_t(x,z,0) = G(x,z)$. Because $f_n$ and
$k_n$ are even over $v$, we have
\begin{equation}
	\label{eqn:decomp}
	F(x,z) = \sum_{n=-\infty}^\infty \int_{c_1}^{c_2} (\alpha_n(v) + \alpha_n(-v))f_n(z; v)e^{ik_n(v)x} \diff{v}.
\end{equation}
We can write this as an inverse Fourier transform by substituting $w=k_n(v)$.
We observe that the bounds on each integral now depend on $n$ as shown in the
following. Let
$b_n = \left.\abs{n}\pi c_1 \middle/ L\sqrt{c_2^2-c_1^2}\right.$, then
\begin{align*}
	F(x,z) = -\sum_{n=0}^\infty \int_{b_n}^\infty \frac{[\alpha_n(k_n^{-1}(w)) + \alpha_n(-k_n^{-1}(w))]}{k_n'(k_n^{-1}(w))} f_n(z; k_n^{-1}(w))e^{iwx} \diff{w} \\
	+ \sum_{n=0}^\infty \int_{-\infty}^{-b_n} \frac{[\alpha_{-n}(k_{-n}^{-1}(w)) + \alpha_{-n}(-k_{-n}^{-1}(w))]}{k_{-n}'(k_{-n}^{-1}(w))} f_n(z; k_{-n}^{-1}(w))e^{iwx} \diff{w}.
\end{align*}
We can significantly simplify this integral by defining the sequence
$A_0,A_1,\dotsc$ of piecewise functions,
\[
	A_n(w)=\begin{cases}
		\alpha_n(k_n^{-1}(w)) + \alpha_n(-k_n^{-1}(w))						 & \text{if } w \geq  b_n, \\
		\alpha_{-n}(k_{-n}^{-1}(w)) + \alpha_{-n}(-k_{-n}^{-1}(w)) & \text{if } w \leq -b_n, \\
		0                                                          & \text{otherwise.}
	\end{cases}
\]
Further, $k_{-n}(v) = -k_n(v)$, so we have $k_{-n}^{-1}(w) = k_{n}^{-1}(-w)$. In total,
\[
	F(x,z) = - \int_{-\infty}^\infty e^{iwx} \sum_{n=0}^\infty \frac{A_n(w)}{k_n'(k_n^{-1}(\abs{w}))} f_n(z; k_n^{-1}(\abs{w})) \diff{w},
\]
and applying the inverse Fourier transform to both sides results in
\[
	\frac{1}{2\pi}\int_{-\infty}^\infty F(x,z) e^{-iwx} \diff{x} = -\sum_{n=0}^\infty \frac{A_n(w)}{k_n'(k_n^{-1}(\abs{w}))} f_n(z; k_n^{-1}(\abs{w})).
\]

The eigenfunctions from \eqref{eqn:loveBVP} are orthogonal with respect to the
weight function $r(z;v) = 1-(v/c(z))^2$. They satisfy the equation
\[
	\int_{-\infty}^L f_n(z; v)f_m(z; v)r(z; v) \mathrm{d}z =
	\begin{cases}
		-\frac{L}{2}\Omega_1(v)^2 & \text{if } \abs{m} = \abs{n}, \\
		0 & \text{otherwise,}
	\end{cases}
\]
where the minus sign appears in the $r-$norm because $r(z;v) < 0$ for $0\leq z < L$.
With this, we isolate for each term in the summation above by multiplying each
side by $f_n(z;k_n^{-1}(\abs{w}))r(z,k_n^{-1}(\abs{w}))$ and integrating over
$z$. The resulting formula for each $A_n(w)$ is,
\[
	A_n(w) = \frac{k_n'(k_n^{-1}(\abs{w}))}{\pi L\Omega_1(k_n^{-1}(\abs{w}))^2}\int_{-\infty}^L\int_{-\infty}^\infty F(x,z) f_n(z; k_n^{-1}(\abs{w}))r(z; k_n^{-1}(\abs{w})) e^{-iwx}\diff{x}\diff{z}
\]
With this, we are halfway to solving for each $\alpha_n(v)$ because,
\begin{align*}
	A_n(\pm k_n(v)) &= \frac{k_n'(v)}{\pi L\Omega_1(v)^2}\int_{-\infty}^L\int_{-\infty}^\infty F(x,z) f_n(z;v)r(z;v) e^{\mp ik_n(v)x}\diff{x}\diff{z} \\
							&= \alpha_{\pm n}(v) + \alpha_{\pm n}(-v), \quad n\geq0.
\end{align*}
With the initial condition for $u_t(x,z,0) = G(x,z)$ and the same logic as
presented above, we compute a similar formula for the difference between two coefficients,
\begin{align*}
	 \pm(\alpha_{\pm n}(v) - \alpha_{\pm n}(-v)) = \frac{ik_n'(v)}{\pi L\Omega_1(v)^2 v k_n(v)}\int_{-\infty}^L\int_{-\infty}^\infty G(x,z) f_n(z;v)r(z;v) e^{\mp ik_n(v)x}\diff{x}\diff{z}.
\end{align*}

For our purposes, we only consider $u_t(x,z,0) = 0$, so
$\alpha_n(v) = \alpha_n(-v)$, $\alpha_{-n}(v) = \overline{\alpha_n(v)}$, and
the solution takes the form
\[
	u(x,z,t) = 4\sum_{n=0}^\infty \int_{c_1}^{c_2} f_n(z;v) \cos(k_n(v)vt) \Re\left(\alpha_n(v)e^{ik_n(v)x}\right)\diff{v}.
\]
Implicit in the assumption \eqref{eqn:linearlove_ansatz} is that the phase does
not change with depth $z$. This assumption is carried through to Equation
\eqref{eqn:total_linear_solution}. As a consequence of this assumption and the
fact that the Sturm-Liouville problem \eqref{eqn:loveBVP} is not regular, the
set of eigenfunctions $\{f_n\}$ is not complete. The decomposition
\eqref{eqn:decomp} holds only for a subset of initial conditions of the
problem for \eqref{eqn:linearlove}.

%

In Section \ref{sec:numerical_linear_love}, we use the method of lines to
numerically study the initial behaviour of linear and nonlinear Love waves
subject to a sudden explosion. While doing so, we record the velocity of the
waves travelling horizontally along the surface and interface. We find that
these velocities initially differ, but eventually take on the same value. As
such, the solution we derived can describe sufficiently large-time behaviour of
a Love wave moving along the $x$ direction.

\section{Nonlinear hyperelasticity and constitutive models}
\label{sec:hypervisco}

The hyperelasticity framework provides a natural generalization of linear waves
corresponding to small displacements in nonlinear materials, including
geological formations, to the case where displacements are not small. In this
section, we describe the behaviour of
incompressible hyperelastic media undergoing finite (possibly large) strains
when subjected to external stress. We briefly recall the
notation and the main ingredients of mathematical models in incompressible
hyperelasticity required for this study (see, e.g., Refs.~\cite{ciarlet1988mathematical, Marsd, Bower, thesisBader} for full details).
Vectors and tensors are indicated with boldface characters and their entries
are written in italics. Entries of vectors and tensors are written in italics,
and summation in repeated indices is assumed where appropriate. We use
Cartesian coordinates and the flat space metric $g^{ij}=\delta^{ij}$.

\subsection{Hyperelastic materials: equations of motion}
\begin{figure}[htbp]
\centering
\fbox{
\includegraphics[width=0.6\linewidth]{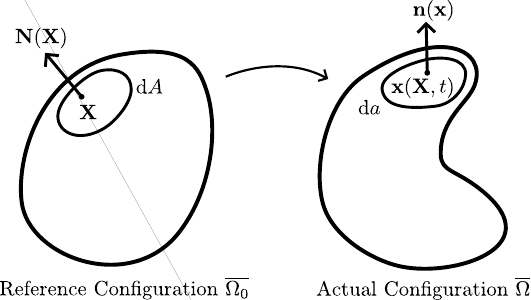}
}

\footnotesize
\caption{A general illustration of the deformation $\mathbf{x}(\mathbf{X},t)$
acting on the reference configuration after some time $t$.}
\label{fig:deform}
\end{figure}

Consider an elastic solid parameterized by material (Lagrangian) coordinates
$\vec{X}=(X,Y,Z)=(X^1,X^2,X^3)$ in a spatial region $\overline{\Omega}_0\subset
\mathbb{R}^3$ with a piecewise-smooth boundary. We call $\overline{\Omega}_0$
the reference configuration or the natural state of the solid. After
deformations, the actual shape of the solid at time $t$ is given by the
time-dependent Eulerian coordinates $\vec{x}=(x,y,z)=(x^1,x^2,x^3)$ in the
actual configuration $\overline{\Omega}\subset\mathbb{R}^3$. The Eulerian
coordinates can be interpreted as an invertible, smooth, orientation-preserving
map deforming the body over time,
\begin{equation}
	\label{eq:Matcoord}
	\vec{x}=\vec{x} \left(\vec{X},t\right)=\vec{X}+\vec{u}\left(\vec{X},t\right).
\end{equation}
The map $\vec{x}$ is also referred to as a deformation field.
$\vec{u}=\vec{u}(\vec{X},t)$ has the meaning of displacement and is
not assumed to be small. Figure \ref{fig:deform} illustrates this situation
along with area elements, the normal vector $\vec{N}$ about a point
$\vec{X}$ in the reference configuration, and its corresponding normal
vector $\vec{n}$ point $\vec{x}$ in the actual configuration. The velocity and the
acceleration of a material point $\vec{X}$ are given by $\vec{v}(\vec{X},t) =
\pdv*{\vec{x}}{t}$ and $\vec{a}(\vec{X},t) = \pdv*[2]{\vec{x}}{t}$. We aim to
describe a system of PDEs describing the deformation field $\mathbf{x}$ in
terms of its derivatives in $\vec{X}$ and $t$.


The deformation gradient $\tens{F}$ is the Jacobian of $\vec{x}$ with
respect to $\vec{X}$,
\begin{equation}
	\label{eq:ch15a}
	\vec{F}(\vec{X},t) = \mathrm{grad}_{\vec{X}} \vec{x}, \qquad
	F_j^i = \frac{\partial x^i}{\partial X^j}.
\end{equation}
Because $\vec{x}$ is an orientation-preserving map, $J\equiv\det(\tens{F})>0$.
The map $\vec{x}$ preserves volumes for incompressible materials, so $J=1$.
We note that when reasonable, we write formulae in terms of general $J$, and
then substitute $J=1$. If the density of the elastic material in the Lagrange
configuration is represented by $\rho_0=\rho_0(\vec{X})$, the current
time-dependent mass density in Lagrangian coordinates takes the form
\[
	\rho(\vec{X},t) = \rho_0(\vec{X})/J = \rho_0(\vec{X}).
\]



The left and right Cauchy-Green deformation tensors $\tens{B} =
\tens{F}\tens{F}^T$ and $\vec{C} = \tens{F}^T\tens{F}$ play an essential role
in the formulation of the equations of motion in solid mechanics.
For our purposes, we regard $\tens{B}$ and $\vec{C}$ as matrices to avoid
unnecessary tensor algebra. $\tens{B}$ and $\tens{C}$ are symmetric
positive-definite and share the same set of eigenvalues. The square roots of
these eigenvalues determine the singular values
$\left(\sigma_1,\sigma_2,\sigma_3\right)$ of $\tens{F}$ which are referred to as the
principal stretches.

Let $\vec{t}$ be the traction vector representing force per unit area acting on
the area element $\mathrm{d}a$ with unit normal vector $\vec{n}$ in the actual
configuration. According to the Cauchy theorem, $\vec{t}=\tens{\sigma}\vec{n}$,
where $\tens{\sigma}=\tens{\sigma}\left(\vec{x},t\right)$ is the symmetric
Cauchy stress tensor. Let $\vec{T}$ be the corresponding traction vector in the
reference configuration acting on the undeformed area element $\mathrm{d}A$
with unit normal vector $\vec{N}$. $\vec{T}$ is given by
$\vec{T}=\tens{P}\tens{N}$, where $\tens{P}=\tens{P}\left(\vec{X},t\right)$
denotes the asymmetric first Piola-Kirchhoff stress tensor, related to the
Cauchy stress through $\tens{P}=J\tens{\sigma}\tens{F}^{-T}$. Here
$\tens{F}^{-T}$ is the inverse of the transpose of $\tens{F}$. The equations of
motion of $\vec{x}$ are given directly in terms of $\tens{P}$.  The second
Piola-Kirchhoff stress tensor $\tens{S}$ is given in terms of the deformation
gradient $\tens{F}$ and the first Piola-Kirchhoff stress tensor $\tens{P}$ by
$\tens{S}=\tens{F}^{-1}\tens{P}$.

The stress tensors $\tens{P}$ and $\tens{S}$ can be explicitly determined by assuming the
form of $\tens{x}$ (e.g. through specification of variables $X^1,X^2,$ or $X^3$ or their combinations
that $x^1,x^2,$ or $x^3$ depend on) and a scalar-valued volumetric
hyperelastic strain energy density $W^H=W^H(\vec{X},\tens{F})$ (sometimes
called stored energy density function). In the incompressible hyperelasticity
framework, the Piola-Kirchhoff stress tensors are determined by the strain
energy density through the formulas
\begin{equation}
	\label{eq:PCformulas}
	\tens{P} = -p\tens{F}^{-T} +  \rho_0 \pdv{W^H}{\tens{F}} = \tens{F}\tens{S},\qquad
	\tens{S} = -p\tens{C}^{-1} + 2\rho_0 \pdv{W^H}{\tens{C}},
\end{equation}
where $p=p(\vec{X},t)$ denotes hydrostatic pressure. Partial derivatives
taken with respect to a matrix/tensor are interpreted as a symmetrized partial
derivative with respect to each entry of the matrix/tensor. For example,
the entries of $\pdv*{W^H}{\tens{C}}$ are given by
\begin{equation}
	\label{eq:19e1}
	\left(\pdv{W^H}{\tens{C}}\right)_{ij} = \frac{1}{2}\left( \pdv{W^H}{C^{ij}} + \pdv{W^H}{C^{ji}} \right).
\end{equation}

For isotropic, homogenous, frame-indifferent, hyperelastic media, the strain
energy density function can be written as a function of the three principal
invariants $I_1$, $I_2$, and $I_3$ of the Cauchy-Green deformation tensors
$\tens{B}$ and $\tens{C}$
\begin{equation}
	\label{eq:19a}
	W^H = W^H_\mathrm{iso}(I_1,I_2,I_3).
\end{equation}
The principal invariants are the coefficients of the characteristic polynomial
of $\tens{B}$,
\begin{equation}
	\label{eq:19g}
	I_1 = \trace(\tens{B}), \quad
	I_2 = \frac{1}{2}\left(I_1^2-\tr(\vec{B}^2)\right), \quad
	I_3 = \det(\tens{B}) = J^2.
\end{equation}
Generally, a set of invariants can be any independent combination of principal
stretches (meaning they cannot be written in terms of the other invariants).
Because $\vec{x}$ preserves volumes (i.e. $I_3=J^2=1$), the strain energy
density has a simpler form $W^H=W^H(I_1,I_2)$. We note that when modelling
anisotropic materials, $W^H$ additionally depends on direction of anisotropy,
such as the direction of embedded fibers, through additional invariants (see,
e.g., \cite{ciarlet1988mathematical,Marsd,cheviakov2016one} and references
therein).

The formulae for $W^H$ only involve its partial derivatives, so we may choose
the constant term. Observe that in the natural state $\vec{x}=\vec{X}$, the
deformation gradient $\tens{F} = \tens{I}$ is the identity matrix, and
$I_1=I_2=n$ in $n$ dimensions. In three dimensions, it is common to take
\begin{equation}
	\label{eq:W:inc:3D}
	W^H = W^H(I_1-3, I_2-3).
\end{equation}
We may then take $W^H(0,0) = 0$ in the natural state.

The following are examples of strain energy density functions. Mooney-Rivlin
solids obey the following strain energy density function
\begin{equation}
	\label{eq:19j}
	W^H = a(I_1-3)+b(I_2-3),
\end{equation}
where $a,b\geq0$ are constant parameters. Any incompressible, lubber-like
material is a Mooney-Rivlin solid. For $b=0$, \eqref{eq:19j} reduces to the
strain energy density function of a neo-Hookean material
\begin{equation*}
	W^H = a(I_1-3).
\end{equation*}
The linear wave equation is the equation of motion for shear waves travelling
through a neo-Hookean solid. There are generalized version of both the
Mooney-Rivlin and neo-Hookean models. The generalized Mooney-Rivlin function is
a bivariate polynomial in $I_1-3$ and $I_2-3$,
\begin{equation}
	\label{eq:20a11}
	W^H = \sum_{i=0}^k \sum_{j=0}^l c_{ij}(I_1-3)^i(I_2-3)^j,
\end{equation}
where $c_{ij}$ are constant material coefficients with $c_{00}=0$. These model
what are called polynomial-type materials \cite{hartmann2001parameter}. One
such generalized Mooney-Rivlin function is the Murnaghan potential. In a
two-dimensional setting, the Murnaghan potential has been recently employed in
Ref.~\cite{nobili2024weakly} to derive a weakly nonlinear analog of the
\emph{Love hypothesis} (which relates the radial displacement and the
longitudinal strain) for a wave propagation model in an elastic rod undergoing
shear and longitudinal axially symmetric displacements.

The generalized neo-Hookean model takes the strain energy density function to
be any function of $I_1-3$ without a dependence on $I_2-3$,
\begin{equation*}
	W^H = W^H(I_1-3).
\end{equation*}
Much of the following is restricted to generalized neo-Hookean materials
because allowing $W^H$ to depend on $I_2$ significantly complicates the
equations of motion. We apply the results of the following section to a Yeoh
model \cite{yeoh1993some}
\begin{equation}
	\label{eq:yeoh}
	W^H = \sum_{i=0}^k c_i(I_1-3)^i,
\end{equation}
with $k=3$, resulting in the polynomial
\begin{equation}\label{eq:n1}
	W^H = \mu(I_1-3) + \frac{1}{2}M(I_1-3)^2 + \frac{1}{3}A(I_1-3)^3,
\end{equation}
where $\mu,M,A$ are constants. For theoretical reasons explained shortly, it is
also desireable for $W^H$ to be a convex polynomial. The coefficients of the
above polynomial models can be fit to a given material's stress-strain curve.
Further generalizations to the above strain energy density functions are
possible and can be found in, e.g., \cite{AgyThesis2020,
cheviakov2012symmetry}. These are useful if a material's stress-strain curve
does not accurately fit to the above forms.

It is desirable for the strain energy density function to be \emph{polyconvex}.
This is the case for generalized Mooney-Rivlin functions (and their
specializations) whenever $c_{ij}\geq0$ for each $i,j$
\cite{ciarlet1988mathematical}. Models built from a polyconvex strain energy
density function have the following three desirable properties. First,
increasing the strain increases the elastic energy. Second, the natural state
corresponds to a stable equilibrium. Lastly, polyconvexity guarantees the
existence of solutions of boundary value problems in linear elasticity
\cite{kambouchev2007polyconvex}. Many realistic materials possess the first two
properties, so they should be modelled by polyconvex strain energy density
functions. The last property is desirable in practice.

To provide the mathematical definition of polyconvexity, a real valued function
$G=G(\tens{F})$ of $3$ by $3$ tensors is polyconvex if it can be written as
\[
	G(\tens{F}) = g(\tens{F},\mathrm{cof}\tens{F},\det\tens{F})
\]
where $g$ is a (typically nonunique) convex function of the entries of
$\tens{F}$, $\tens{F}$'s matrix of cofactors $\mathrm{cof}\tens{F}$, and
$\det\tens{F}$ \cite{Marsd}. For the examples, suppose
$h\colon\mathbb{R}\to\mathbb{R}$ is convex. Any function of the form
$G(\tens{F})=h(\det\tens{F})$ is polyconvex but not necessarily convex with
respect to the entries of $\tens{F}$. Further, $G(\tens{F})=h(|\tens{F}|)$
where $|\tens{F}|$ denotes the Frobenius norm of $\tens{F}$ is both polyconvex
and convex with respect to the entries of $\tens{F}$.



The full system of equations of motion for incompressible hyperelastic
materials in three dimensions is given by
\begin{subequations}
\label{eq:chse5}
\begin{align}
	\label{eq:ch2282}
	\rho_0\pdv[2]{\vec{x}}{t}=\mathrm{div}_{\vec{X}}\tens{P} + \vec{D}_f,\\[2ex]
	\label{eq:ch2292}
 	\tens{P}\tens{F}^T=\tens{F}\tens{P}^T, \\[2ex]
  \label{eq:ch22312}
 	J=\det(\tens{F})=1,
\end{align}
\end{subequations}
where $\vec{D}_f$ are the body forces per unit volume (e.g. gravity). The
$i^{\mathrm{th}}$ component of the divergence of the tensor $\tens{P}$ with
respect to the Lagrangian coordinates $\vec{X}$ is the divergence of the
$i^{\mathrm{th}}$ row of $\tens{P}$. So,
\[
	\left(\div_{\vec{X}}\tens{P}\right)^i = \div_{\vec{X}}\left(\tens{P}^i\right)
	= \pdv{}{X^j}P^i_j.
\]
The equations of motion are a result of three practical balance laws
(conservation of mass and angular momentum, and balance of linear momentum).
The balance law \eqref{eq:ch2282} is known as the momentum equation. Equation
\eqref{eq:ch2282} expresses a conservation law if $\vec{D}_f=0$, where it
becomes a total divergence. The condition \eqref{eq:ch2292} expresses
conservation of angular momentum and is equivalent to requiring the Cauchy
stress tensor be symmetric $\tens{\sigma}=\tens{\sigma}^T$. For isotropic
materials, as well as in some other cases, this symmetry condition is
identically satisfied (e.g., Ref.~\cite{cheviakov2012symmetry}).

To summarize (and provide a preview of the following), one can use equation
\ref{eq:chse5} to model an incompressible hyperelastic material after assuming
the form of $\vec{x}$, fitting a strain energy density function $W^H$ to their
material's stress-strain curve, and then finding the first Piola-Kirchhoff
stress tensor.
This process is made easier when the material is isotropic, homogenous and
frame-indifferent because then $W^H$ is a function of three arguments (the
principal invariants of $\tens{B}=\tens{F}\tens{F}^T$). In that case, one can
assume a general form for $W^H$, then the PDE \eqref{eq:ch2282} will model any
material whose stress-strain curve is well approximated by such $W^H$.

\section{Nonlinear hyperelastic shear waves}
\label{Sec:NonlinearHyperelasticLoveWaves}

\subsection{The general wave equation for an arbitrary strain energy density function}
\label{sec:nonlWderivation}

Consider a shear wave with displacements in the $Y-$direction propagating
through the $X-Z$ plane (which is antiplane motion) described by
\begin{equation}
	\label{eq:2ns11}
	\vec{x} =
	\begin{bmatrix}
		x \\ y \\ z
	\end{bmatrix}
	=
	\begin{bmatrix}
		X \\ Y \\ Z
	\end{bmatrix}
	+
	\begin{bmatrix}
		0 \\ v(t,X,Z) \\ 0
	\end{bmatrix}.
\end{equation}
Our aim is to derive the equations of motion for $v(X,Z,t)$. As discussed in
Section \ref{sec:hypervisco}, these follow from computing $\tens{P}$. The
deformation gradient and its inverse are
\begin{equation}
	\label{eq:2ss2}
	\vec{F} = \begin{bmatrix}
		1   & 0 & 0   \\
		v_X & 1 & v_Z \\
		0   & 0 & 1
	\end{bmatrix},
	\quad
	\vec{F}^{-1}=\begin{bmatrix}
		1   & 0 &  0   \\
 	 -v_X & 1 & -v_Z \\
 		0   & 0 &  1
	\end{bmatrix},
\end{equation}
where $v_X=\pdv*{v}{X}$, and $v_Z=\pdv*{v}{Z}$ represent the amount of shear.
The left and right Cauchy-Green deformation tensors $\tens{B}$ and $\vec{C}$
are
\begin{equation}
	\label{eq:2ss3}
	\vec{C}=\begin{bmatrix}
		1+v_X^2 & v_X & v_Xv_Z \\
		v_X     & 1   & v_Z    \\
		v_Xv_Z  & v_Z & 1+v_Z^2
	\end{bmatrix}, \quad
	\vec{B}=\begin{bmatrix}
		1   & v_X           & 0   \\
		v_X & v_X^2+v_Z^2+1 & v_Z \\
		0   & v_Z           & 1
	\end{bmatrix}.
\end{equation}
The principal invariants for the deformation field \eqref{eq:2ns11} are
\[
	I_1 = I_2 = 3 + v_X^2 + v_Z^2 = 3 + \norm{\mathrm{grad}\,v}^2 = |\tens{F}|^2, \quad I_3=1.
\]
where $|\tens{F}|$ is the Frobenius norm of $\tens{F}$. The shear
displacements \eqref{eq:2ns11} are naturally incompressible because
$I_3=\det(\vec{F})=1$. It is important to note that, although $I_1=I_2$, we
cannot use this fact until after expanding \eqref{eq:PCformulas} and
\eqref{eq:19e1}. Doing otherwise would miss a necessary condition to ensure the
pressure is well defined.


Equation \eqref{eq:PCformulas} yields the first Piola-Kirchhoff stress tensor
\begin{equation}
	\label{eq:cssa71}
	\tens{P} = -p\tens{F}^{-T} + 2W^H_1\tens{F} - 2W^H_2\tens{B}^{-1}\tens{F}^{-T},
\end{equation}
where $W^H=W^H(I_1,I_2)$, $W^H_1\equiv\pdv{W^H}{I_1}$, $W^H_2\equiv\pdv{W^H}{I_2}$, and $p=p(X,Z,t)$.
Substituting \eqref{eq:cssa71} into the equations of motion \eqref{eq:ch2282}
yield the following system of PDEs
\begin{subequations}
	\label{eq:shear:genW}
	\begin{gather}
		 q_X+2(W^H_2v_X^2)_X+2(W^H_2v_Xv_Z)_Z=0, \qquad q_Z+2(W^H_2v_Xv_Z)_X+2(W^H_2v_Z^2)_Z=0, \label{eq:shear:genW:x:z} \\[1ex]
		 \rho_0 v_{tt} = 2\left( (v_X(W^H_1+W^H_2))_X + (v_Z(W^H_1+W^H_2))_Z \right), \label{eq:shear:genW:y}
	\end{gather}
\end{subequations}
where $q=p-2W^H_1-2(v_X^2+u_Z^2+2)W^H_2$. Cross differentiating the two
equations for $q$ in \eqref{eq:shear:genW:x:z} yields the compatibility
condition
\begin{equation}
	\label{eq:compatibility}
	\left((W^H_2v_Xv_Z)_X+(W^H_2v_Z^2)_Z\right)_X - \left((W^H_2v_X^2)_X+(W^H_2v_Xv_Z)_Z\right)_Z=0.
\end{equation}
This condition can be written in the language of vector calculus as
\[\mathrm{curl}\left(\div\left(W_2(||\grad{v}||^2)\grad{v}\otimes\grad{v}\right)\right)=0.\]

The system \eqref{eq:shear:genW} and \eqref{eq:compatibility} is
overdetermined. Such systems do not typically have solutions, although the
problem's nonlinearity makes it difficult to determine existence or uniqueness
of solutions with confidence. We also note that, because
$I_1=u_X^2+u_Z^2+3=I_2$, there are many different forms of $W^H$ resulting in
the same equation \eqref{eq:shear:genW:y}. The strain in a material undergoing
a displacement of the form \eqref{eq:2ns11} is a univariate function of the
stress because one can only observe the values of
\begin{equation}
	\label{eq:NeqSED}
	W(I_1) = W^H(I_1,I_1).
\end{equation}
Fitting $W$ to this material's stress-strain curve provides no information on
$W^H_1$ or $W^H_2$ individually, but entirely determines the equation
\eqref{eq:shear:genW:y} as $W'(I_1)=W^H_1(I_1,I_1)+W^H_2(I_1,I_1)$. We avoid
the intricacies of \eqref{eq:compatibility} by restricting our attention to
generalized neo-Hookean materials, where $W^H_2=0$. This restricts the
resulting model's applicability in practice, because there are materials that
necessarily depend on $I_2$. The issue of determining other conditions on $W^H$
such that $W^H_2\neq0$ and the compatibility condition \eqref{eq:compatibility}
holds for all solutions to \eqref{eq:shear:genW:y} has been explored in
\cite{Saccomandi}. This issue is outside the scope of the current work.

Restricting our attention to generalized neo-Hookean materials also makes it
easier to determine whether a strain energy density function is polyconvex. In
\cite[Sec.~2.2]{Suchocki2021}, Suchocki \emph{et al.} state that such
$W^H(I_1)$ is polyconvex if and only if $W^H(I_1)$ is convex. As such, we need
only enforce $W'' > 0$.


For generalized neo-Hookean materials (i.e. $W^H_2=0$), the equations for the
hydrostatic pressure reduce to,
\[
	p_X=2(W'(v_X^2+v_Z^2))_X, \quad p_Z=2(W'(v_X^2+v_Z^2))_Z,
\]
yielding the solution in terms of $v_X$ and $v_Z$
\begin{equation}
	\label{eq:gen:nonl:P}
	p = 2W'(v_X^2+v_Z^2) + p_0(t),
\end{equation}
where $p_0(t)$ is the ambient hydrostatic pressure. If the gravity force
$\vec{e}=\rho_0g\vec{e}_Z$ is taken into account, equation \eqref{eq:gen:nonl:P}
is replaced by $p=2W'(v_X^2+v_Z^2)+p_0(t)+\rho_0gZ$, with $Z$ directed downward.

\begin{theoremp}
The $Y-$displacement $v$ of a shear wave given in equation \eqref{eq:2ns11}
travelling through a generalized neo-Hookean material ($W^H_2=0$ so
$W=W^H(I_1)$) satisfies the nonlinear wave equation \eqref{eq:shear:genW:y},
explicitly written as
\begin{equation}
	\label{eq:shear:genW:explicit}
	\rho_0v_{tt} = 2W'(v_X^2+v_Z^2)(v_{XX}+v_{ZZ}) + 4W''(v_X^2+v_Z^2)(v_X^2v_{XX} + 2v_Xv_Zv_{XZ} + v_Z^2v_{ZZ}).
\end{equation}
\end{theoremp}

We refer to equation \eqref{eq:shear:genW:y}, \eqref{eq:shear:genW:explicit} as
the \emph{general nonlinear shear wave equation}. The PDE
\eqref{eq:shear:genW:explicit} is linear if and only if $W'' \equiv
\pdv*[2]{W^H(I_1)}{I_1}=0$ (i.e. $W$ is linear in its single argument
$I_1-3$). This is the case for Mooney-Rivlin materials with strain energy
density function $W^H=a(I_1-3)$. In this case, the Young's modulus is $a$, the
wave equation (\ref{eq:shear:genW:explicit}) reduces to the linear PDE
$\rho_0v_{tt}=2a\left(v_{XX}+v_{ZZ}\right)$, and the hydrostatic pressure is
$p=2a+\rho_0gZ$.

\subsection{The general nonlinear shear wave equation for the cubic Yeoh model}
\label{sec:nonlW:Murnaghan}

Substituting the cubic Yeoh model \eqref{eq:n1} into the general nonlinear
shear wave equation \eqref{eq:shear:genW:explicit} yields
\begin{subequations}\label{eq:y1M}
\begin{equation}
	\label{Y1M6}
	\begin{split}
		\rho_0v_{tt}=2\left(\mu+M\left( v_X^2+v_Z^2 \right)+A\left(v^{2}_{X}+v^{2}_{Z}\right)^{2}\right)(v_{XX}+v_{ZZ})\\
				\qquad+4\left(M + 2A(v_X^2+v_Z^2)\right)(v_X^2v_{XX}+2v_Xv_Zv_{XZ}+v_Z^2v_{ZZ}).
	\end{split}
\end{equation}
The hydrostatic pressure \eqref{eq:gen:nonl:P} in this case is given by
\begin{equation}
	\label{Y1M7}
	p = 2\left(\mu + M(v_X^2+v_Z^2) + A(v_X^2 + v_Z^2)^2 \right) + p_0(t)+\rho_0gZ.
\end{equation}
\end{subequations}
Equations \eqref{eq:y1M} contain nonlinear polynomial terms of
the third and fifth orders in terms of $v$. Cubic effects are controlled by
the coefficient $M$ and quintic effects by the
coefficient $A$.

Writing equation \eqref{Y1M6} with gradients $\grad = \mathrm{grad}_{(X,Z)}$
makes it more apparent what effects each parameter has and that this equation
is a total divergence,
\begin{equation}\label{eqn:nablas}
	\frac{\rho_0}{2} v_{tt} = \div\left( \left[\mu + M\norm{\grad v}^2 + A\norm{\grad v}^4\right]\grad v \right).
\end{equation}
The expression in square brackets is $W_1$ composed with $\norm{\grad v}^2$.

To our knowledge, the PDEs \eqref{eq:y1M}, \eqref{eqn:nablas} have not
previously appeared in the literature. Similar models of nonlinear shear waves
based on displacements (\ref{eq:2ns11}) were considered in Kalyanasundaram
\cite{kalyansaundaram1981finite}, Teymur \cite{teymur1988nonlinear}, and
Rushchitsky \cite{rushchitsky2013nonlinear,rushchitsky2014nonlinear}, who
assumed a compressible hyperelastic material, and employed a different
(Murnaghan) hyperelastic potential. In our approach, because the deformation
field \eqref{eq:2ns11} naturally preserves volumes, new nonlinear wave
equations are coupled to the hydrostatic pressure fields that depend on
displacements according to \eqref{Y1M7}.

We note that because the nonlinearity in the PDE \eqref{eqn:nablas} involves
only gradients, the dispersion relation for perturbations $v = v_0 + \epsilon
\exp(i(k_1 X_1 + k_3 X_3 - \omega t))$ around a constant equilibrium $v_0$
coincides with the linear wave equation dispersion relation
\[
	\omega^2=\dfrac{2\mu}{\rho_0}(k_1^2+k_3^2).
\]
However, if the equation \eqref{eqn:nablas} is linearized about a uniform shear
solution $v = v_0 + sX_1$, $s=\const$, that is, in the ansatz
$v = v_0 + sX_1 + \epsilon \exp(i(k_1 X_1 + k_3 X_3 - \omega t))$, the dispersion relation takes
the form
\[
	\omega^2=\dfrac{2}{\rho_0}\left(\mu(k_1^2+k_3^2) + Ms^2(3k_1^2+k_3^2) + As^4(5k_1^2+k_3^2) \right).
\]

\section{Nonlinear Viscoelastic shear waves} \label{sec:NonlinearVisco}

In this section, we add non-elastic effects responsible for \emph{energy
dissipation} to our model. Considering both elasticity and viscosity ensures
our model of large displacements in nonlinear materials remain realistic over
long periods of time. Materials such as geological formations, composite
materials, and biological tissues exhibit viscoelastic properties.  Various
approaches exist for the mathematical description of viscoelasticity, including
rational and irreversible thermodynamics, finite viscoelasticity, and
hyper-viscoelasticity. For a more detailed review, see
Refs.~\cite{thesisBader,cheviakov2016one}. In the current work, we use the
hyper-viscoelasticity framework \cite{holzapfel2000nonlinear}. This framework
uses both a hyperelastic strain energy density function $W^H$ to describe
elastic effects, and a dissipative potential $W^V$ to describe viscous
phenomena. Hyper-viscoelastic models have also been used in fiber-reinforced
material settings \cite{cheviakov2016one,pioletti2000non}.

\subsection{Hyper-viscoelastic constitutive models}
\label{sec:visco:framew}

In the hyper-viscoelasticity framework, the total stress is represented as a
sum of the elastic and the viscous (dissipative) parts
\begin{equation}
	\label{1}
	W = W^H + W^V.
\end{equation}
$W^V$ is generally a function of $\mathbf{C}$ and $\mathbf{\dot{C}} =
\pdv*{\mathbf{C}}{t}$. For homogenous, isotropic materials, $W^V$ can be
written as a function of up to three independent invariants of $\mathbf{C}$ and
up to seven independent invariants of $\mathbf{\dot{C}}$. Invariants of
$\mathbf{\dot{C}}$ are often called pseudo-invariants and similarly, $W^V$ the
pseudo-strain energy density function. The second Piola-Kirchhoff tensor from
equation \eqref{eq:PCformulas} is modified as follows to include the viscoelastic
stress
\begin{equation}
	\label{eq:PK2:viscoel}
	\tens{S}^V = 2\rho_0 \pdv{W^V}{\tens{\dot{C}}},
\end{equation}
The total stress tensor is
\begin{equation}
	\label{eq:PK2:viscoel:tot}
	\tens{S} = \tens{S}^H + \tens{S}^V = -p\tens{C}^{-1} + 2\rho_0\left(\pdv{W^H}{\tens{C}} + \pdv{W^V}{\tens{\dot{C}}}\right).
\end{equation}
As noted in Section \ref{sec:hypervisco}, partial derivatives with respect to a
matrix/tensor are performed elementwise and symmetrized similar to equation
\eqref{eq:19e1}. The equations of motion of the solid are still given by
\eqref{eq:chse5}, with $\tens{P} = \tens{F}\tens{S}$.

For the present work, we use the potential given by
\begin{equation}
	\label{eq:Wv:our}
	W^V=\frac{\eta}{4}J_2(I_1-3), \quad J_2=\tr(\dot{\tens{C}}\dot{\tens{C}}),
\end{equation}
This convex potential, linear in $J_2$ and $I_1 - 3$, performed well in the
short-time memory benchmarks presented by Pioletti et
al.~\cite{pioletti2000non}. For this benchmark, the authors could accurately
fit the single parameter $\eta$ to the stress-strain curve of several patellar
tendons.

To apply the above method, recall the purely elastic model of a shear wave with
displacements in the $Y-$direction propagating through the $X-Z$ plane
described by \eqref{eq:2ns11}. Viscous effects are added to this setup using
the potential $W^V$ \eqref{eq:Wv:our}. The pseudo-invariant $J_2$ takes the
form
\begin{equation}
	\label{eq:Wv:our:J2} 
	J_2 = 2\left(v_{Xt}^2(2v_X^2 + v_Z^2 + 1) + 2v_{Xt}v_{Zt}v_Xv_Z + v_{Zt}^2(v_X^2 + 2v_Z^2 + 1)\right).
\end{equation}
Using the combined elastic-viscoelastic second Piola-Kirchhoff stress tensor
\eqref{eq:PK2:viscoel:tot} to obtain $\tens{P} = \tens{F}\tens{S}$ (cf.
\eqref{eq:PCformulas}) and substituting the latter in the general equation of
motion \eqref{eq:ch2282} with zero forcing (i.e. $\vec{D}_f = \vec{0}$) yields
the following result.

\begin{theoremp}
The $Y-$displacement $v$ of shear waves \eqref{eq:2ns11} in the viscoelastic
setting with an arbitrary strain energy density function $W$ and the viscous
potential \eqref{eq:Wv:our} satisfies the nonlinear wave equation
\begin{equation}\label{eq:shear:genW:explicit:visc}
	\rho_0 v_{tt}=2\left(\left(v_{X}W_{1}\right)_X+\left(v_{Z}W_{1}\right)_Z\right) + \eta Q[v],
\end{equation}
where $Q[v]$ is the viscosity term given by
\begin{equation}
\label{eq:shear:genW:explicit:visc:Q}
	\begin{array}{ll}
	Q[v] = & \left(v^{2}_{X}v_{Xt}\right)_{X}+\left( v^{2}_{Z}v_{Xt}\right)_{X}+\left( v^{2}_{X}v_{Zt}\right)_{Z}+\left( v^{2}_{Z}v_{Zt}\right)_{Z}\\[1ex]
	& + 2\left(v^{4}_{X}v_{Xt}\right)_X + \left(v^{4}_{Z}v_{Xt}\right)_{X} + \left( v^{4}_{X}v_{Zt}\right)_{Z} +2\left( v^{4}_{Z}v_{Zt}\right)_{Z} \\[1ex]
	& + \left((v^{3}_{Z}v_{X}+v^{3}_Xv_Z)v_{Zt}\right)_X+\left( (v^{3}_{Z}v_{X}+v^{3}_{X}v_{Z})v_{Xt}\right)_{Z}\\[1ex]
	& + 3\left(v^{2}_{Z}v^{2}_{X}v_{Xt}\right)_{X}+ 3\left(v^{2}_{Z}v^{2}_{X}v_{Zt}\right)_{Z},
	\end{array}
\end{equation}
\end{theoremp}
Equation \eqref{eq:shear:genW:explicit:visc} is the generalization of the
nonlinear shear wave equation \eqref{eq:shear:genW:y} onto the viscous case,
with $\eta$ denoting a constant viscosity coefficient. By considering only the
cubic terms in this equation, one can write $Q[u]$ as a total divergence,
\[
	Q[u]\approx \mathrm{div}\left(\norm{\grad v}^2\grad v_t\right).
\]
The quintic terms can also be written as a divergence of some function applied
to $\grad v_t$, but the terms in line $3$ of equation
\eqref{eq:shear:genW:explicit:visc:Q} require swapping the entries of
$\grad v_t$. The pressure terms that follow from $X-$ and $Z-$components of the
equations of motion yield (cf. \eqref{eq:shear:genW:x:z})
\begin{equation}
	\label{eq:shear:genW:explicit:visc:pressure}
	p_X=2(W_1)_X + \eta K_1[v], \qquad p_Z=2(W_1)_Z + \eta K_2[v],
\end{equation}
where $K_1[v]$, $K_2[v]$ are certain expressions omitted here for brevity. In
particular, for the pressure $p$ to be defined, compatibility conditions
$p_{XZ}=p_{ZX}$ must be satisfied.

In principle, the equation for $Q[v]$ could be written in terms of an arbitrary
viscosity potential; however, the expression is too unweildy to include here.
The given viscosity potential is sufficient for our purposes. Readers wishing to
alter our construction by using other viscosity potentials can access our \verb|Maple|
script upon request.



\section{Numerical simulations of hyperelastic and hyper-viscoelastic nonlinear Love wave problems}
\label{sec:numerical}
We now proceed to use the method of lines to numerically study the qualitative
behaviour of Love waves propagating through materials governed by the general
nonlinear shear wave equations \eqref{eq:shear:genW:explicit} and
\eqref{eq:shear:genW:explicit:visc}. First, as a reference, we use the neo-Hookean strain
energy density function $W^H = a(I_1 - 3)$ without viscosity. This
results in the familiar linear Love wave equation \eqref{eqn:linearlove}.
Second, we study a hyperelastic material governed by the cubic Yeoh
model described in Subsection \ref{sec:hypervisco} without viscosity, resulting in
the PDE \eqref{Y1M6}. We finish by studying a hyper-viscoelastic material
governed by the cubic Yeoh potential and the viscosity potential from
equation \eqref{eq:Wv:our} resulting in the PDE
\eqref{eq:shear:genW:explicit:visc}. In every experiment, we only consider up
to cubic terms, neglecting the quintic ones. For Love waves propagating
through hyperelastic materials without viscosity, this is equivalent to taking
$A=0$ in equation \eqref{Y1M6}. After including viscosity, we must additionally
assume $\norm{\grad v}$ is sufficiently small to ensure the quintic
terms are dominated by the lower order (linear and cubic) terms.

Similar to Section \ref{sec:linearlove}, we model an isotropic layer overlaying
an isotropic half-space by considering the coefficient $\mu$ of
$\Delta v = (v_{xx} + v_{xx})$ to be a piecewise constant function of $z$.
Every other parameter is constant across the entire domain. The initial
condition is a sudden ``Gaussian explosion'' centered at a point $(0,z_0)$.
\begin{equation}
	\label{eqn:gaussian_IC}
	v(x,z,0) = A\exp\left(-\frac{x^2+(z-z_0)^2}{r^2}\right), \quad v_t(x,z,0) = 0,
\end{equation}
where the explosion's radius $r$ is small relative to the size of the domain.
We compare the behaviour of each solution when an explosion occurs in the lower
half-space ($z_0<0$) for $c_1<c_2$ and $c_2<c_1$, or the upper layer
($0<z_0<L$) for $c_1<c_2$. We do not show the case when the explosion occurs
above the interface with $c_2<c_1$ because the behaviour is not interesting.

\subsection{Simulation setup and method for numerical solution}

The method of lines transforms the problem of solving a PDE into the problem of
solving a system of coupled initial value ODE problems (IVPs), one for each
point on the discretized spatial domain. The primary advantage of the method of
lines is that there exist fast high-precision methods and their software
implementations for numerically solving the resulting IVPs. For the following,
we use MATLAB's \texttt{ode23}, which combines Runge--Kutta methods of order
$2$ and $3$ to produce an error estimate after each time step. The method of
lines is straightforward to implement for PDEs whose domains simple shapes,
such as a rectangle or circle. One alternative to the method of lines is the
finite element method provided, for example, by MATLAB's PDE Toolbox and
various other free and proprietary software packages. Finite element methods
are better suited for complex domain geometries, but our problem domain is
simple. In this setting, the finite element method is largely equivalent to the
method of lines but requires more work per time step.

For the PDEs \eqref{eqn:linearlove}, \eqref{Y1M6}, and
\eqref{eq:shear:genW:explicit:visc}, we compute an approximate solution
over a discretized rectangular domain of width $W$, height $H>L$, and step size
$h$. Any point in this discretized domain is of the form
\begin{align*}
	(x_j,z_k) &=(-W/2 + jh_x, L - kh_z) \\
						&\in \{-W/2, -W/2 + h_x, \dotsc, W/2-h_x, W/2\}\times\{L, L-h_z, \dotsc, h_z-H, -H\}.
\end{align*}
One can derive an ODE for a matrix $U(t)$ such that its $j,k^{\mathrm{th}}$
entry $U_{j,k}(t)$ approximates $v(x_j,z_k,t)$; this is achieved using finite
differences to approximate spatial derivatives. For example, second-order
finite differences yield the ODEs
\begin{multline*}
	v_{tt}(x_j,z_k,t) \approx c(z_k)^2\bigg( \frac{v(x_j+h_x,z_k,t) + v(x_j-h_x,z_k,t) - 2v(x_j,z_k,t)}{h_x^2} \quad + \\
	\frac{v(x_j,z_k+h,t) + v(x_j,z_k-h,t) - 2v(x_j,z_k,t)}{h_z^2} \bigg)
\end{multline*}
where $c(z) = c_1$ if $z$ is in the upper layer (i.e. for $0<z<L$), and $c(z) =
c_2$ if $z$ is in the lower half-space (i.e. for $L<z$). The corresponding ODE
for the matrix $U(t)$ is
\[
	U''_{j,k} = c(z_k)^2\left(\frac{U_{j+1,k} + U_{j-1,k} - 2U_{j,k}}{h_x^2} + \frac{U_{j,k+1} + U_{j,k-1} - 2U_{j,k}}{h_z^2}\right).
\]
This formula is only valid on the interior of the rectangular domain. The
remaining points are determined from the boundary conditions.
Because the initial condition and each PDE are symmetric about the line $x=0$,
we need only compute $v$ for $x\geq0$.

Our implementation of the method of lines in MATLAB featured second-order
finite differences over a rectilinear (nonuniform) mesh. The mesh was chosen
such that more grid points clustered near $(0,L)$, and this is for two reasons.
First, the displacements $u$ are largest (and hence the nonlinear effects are
most prominent) at the start of the simulation near the source of the explosion
$(0,z_0)$. As the simulation progresses, the wave travels in all directions
with a much smaller amplitude, so it is acceptable to use fewer grid points
there. Further, the wave's behaviour above the interface $z=0$ where it can
interact with both the interface and the surface is of more interest to us than
behaviour below the interface. As such, fewer grid points are used below the
interface.

To construct this nonuniform partition, we use a pair of smooth monotone
functions $\xi(x),\eta(z):[0,1]\to[0,1]$ with fixed points at $0$ and $1$ to
transform the uniform discretized domain into a nonuniform one. The nonuniform
mesh is then given by $\xi_j = W\xi(x_j/W+1/2)-W/2$,
$\eta_k = (L+H)\eta((z_k+H)/(L+H))-H$ and we can solve for the derivatives of
$u$ on the nonuniform mesh with the chain rule. For clarity, let $v_1$ be the
derivative of $v$ with respect to its first argument. The chain rule applied to
$(v(\xi(x),z))_x$ implies
\[
	v_1(\xi_j,z) = \frac{(v(\xi_j,z))_1}{\xi'(x_j)}.
\]
We may approximate $(v(W\xi(x_j/W+1/2)-W/2,z))_1$ with any finite difference
method over the uniform partition $\{x_i\}$ without decreasing that method's
rate of convergence, because $\xi'(x_j)$ is known exactly. For our purposes,
the primary advantage of constructing a nonuniform partition in this way is
that second order centered finite differences over $x_{j-1},x_j,x_{j+1}$ remain
second order accurate with the nonuniform mesh. Formulae for $v$'s other
derivatives follow by the same logic. For both $\xi$ and $\eta$, we use a
quadratic of the form $p(x) = ax^2+(1-a)x$ with $a\in[0,1]$. Observe that such
$p$ is monotone increasing with $p(0)=0$ and $p(1)=1$. The slope of $p$ at $0$
is $1-a$, and increasing $a$ causes more mesh points to cluster about $0$.

%
%

For each of the following simulations, we assume the surface is stress-free, so
$v_z(x,0,t)=0$. To first order, this Neumann boundary condition is equivalent
to requiring $U_{j,0}(t) = U_{j,1}(t)$ for all $j,t$ (but higher-order
approximations would also be reasonable). The other boundaries are modelled as
rigid walls (so $U_{0,k} = U_{\mathrm{end},k} = U_{j,\mathrm{end}} = 0$ for all
$j,k$), but $W,H$ are sufficiently large such that these boundary conditions
are irrelevant to the simulation. We solve the resulting system of ODEs in
MATLAB with the function \texttt{ode23}. This function is chosen because it (1)
somewhat accounts for the stiffness inherent in equations with damping such as
parabolic equations when discretized by the method of lines, (2) does not
evaluate the ODE's Jacobian, and (3) uses vector-level parallelism to compute
the $3$D solution matrix $U$ with repeated matrix addition and scalar
multiplication. The linear wave equation and inviscid equation \eqref{Y1M6} are
likely not stiff because they do not possess any damping terms. The equation
\eqref{eq:shear:genW:explicit:visc} likely is stiff 1. because of the damping
and 2. nonstiff solvers struggle to solve the equation numerically. MATLAB has
various stiff ODE solvers such as \texttt{ode15s} or \texttt{ode23s}, but these
require evaluating the ODE's Jacobian. The Jacobian of our ODE is very large
because we took the mesh as fine as reasonably possible on our machine. For
each simulation, there were about $250^2$ mesh points resulting in
$2\times250^2$ ODEs (one ODE for $v$ and another for $v_t$ for each mesh point).
Each ODE depends on the the $4$ points neighbouring $(x_j,z_k)$.
As such, the Jacobian would be a (sparse) matrix of size
$2\times250^2\times250^2\approx10^{10}$ and contain at least
$2\times4\times250^2\approx10^5$ nonzero entries. Further, the nonlinearity of
this PDE system makes the Jacobian's entries nonconstant and tricky to quickly
compute for each time step. As such, we believe the stiff solvers are likely
not worth the effort and increased memory usage over \texttt{ode23}.

\subsection{Linear shear waves and linear Love waves}
\label{sec:numerical_linear_love}

\begin{figure}[htbp!]
\includegraphics[width=\linewidth]{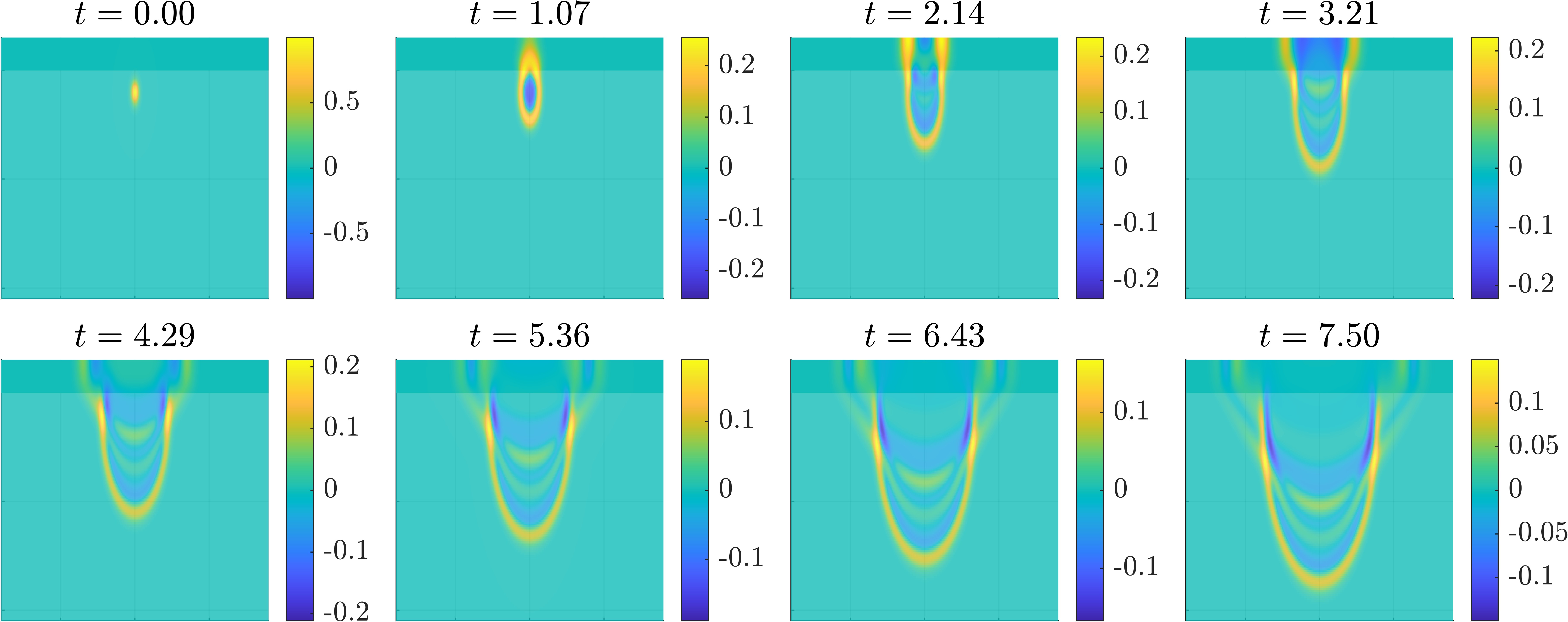}
\footnotesize
\centering

\caption{Plots of a numerical solution $u(x,z,t)$ of equation
	\eqref{eqn:linearlove} with $c_1<c_2$ after a Gaussian explosion centered in
	the lower half-space (i.e. $z_0<0$ in equation \eqref{eqn:gaussian_IC}). In
	this and any similar surface plots in this section, $x$ changes over the
	horizontal axis and $z$ changes over the vertical axis.
	The parameters are $c_1 = 0.1$, $c_2 = 0.2$, $z_0=-0.1$, $r=0.05$, $A=1$,
	$L=0.15$, $W=3.6$, $H=1.8$, and $h = 0.005$.}
\label{fig:gaussian_below_c1c2}
\end{figure}

\begin{figure}[htbp]
\includegraphics[width=\linewidth]{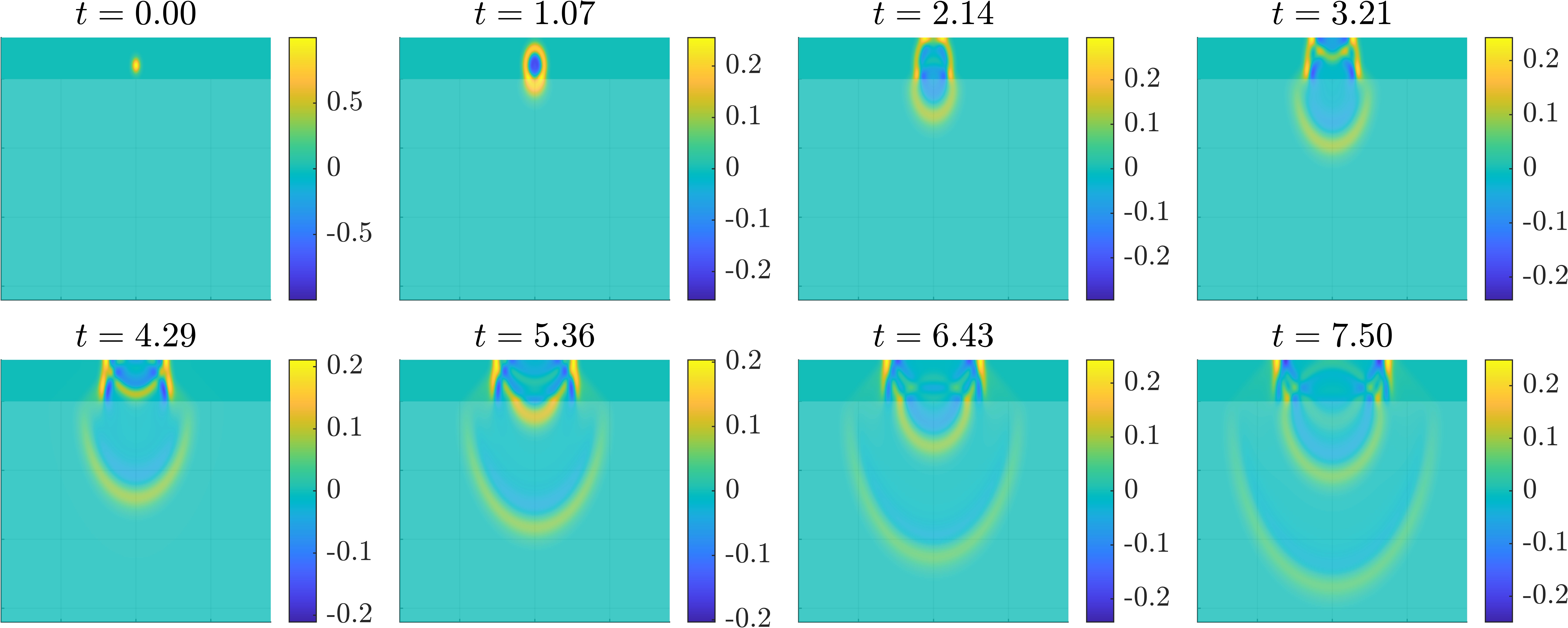}
\footnotesize
\centering

\caption{Plots of a numerical solution of \eqref{eqn:linearlove} with $c_1<c_2$
and $z_0$ in the upper layer. The parameters (bolded to emphasize parameters
that differ from previous figures) are $c_1 = 0.1$, $c_2 = 0.2$,
$z_0=\mathbf{0.1}$, $r=0.05$, $A=1$, $L=\mathbf{0.3}$, $W=3.6$,
$H=\mathbf{1.6}$, and $h = 0.005$.}
\label{fig:gaussian_above_c1c2}
\end{figure}

Figures \ref{fig:gaussian_below_c1c2}, \ref{fig:gaussian_above_c1c2}, and
\ref{fig:gaussian_below_c2c1} contain colour plots of a solution of
the linear Love wave equation \eqref{eqn:linearlove} for various times and
values of $z_0$, $c_1$, and $c_2$. The initial condition is in the top left
corner of each figure, and the interface in each plot is marked by a change
of opacity. We can tell which domain has the faster wave speed based
on how distorted the wave's curvature is. If $c_1 < c_2$ then the outermost
wave propagating below the interface looks qualitatively more like a
semi-circle than if $c_2 < c_1$. Whenever a wave crosses the interface, a
reflection/refraction interaction occurs. So, some proportion of the wave
crossing the interface will reflect and travel in the opposite direction
instead of continually travelling in a straight line. Because the surface is
free of damping, the entire wave is reflected at the surface of the domain.
In particular, it is of interest to observe the case when $c_2<c_1$ (Figure
\ref{fig:gaussian_below_c2c1}), which violates the Love wave existence
condition \eqref{eq:LW:ex:cond}. In the numerical solution, unlike the cases
when $c_1<c_2$ (e.g., Figure \ref{fig:gaussian_below_c1c2}), most of the wave
energy remains below the interface, while the wave above the interface appears
to decay in time.

\begin{figure}[htbp!]
\includegraphics[width=\linewidth]{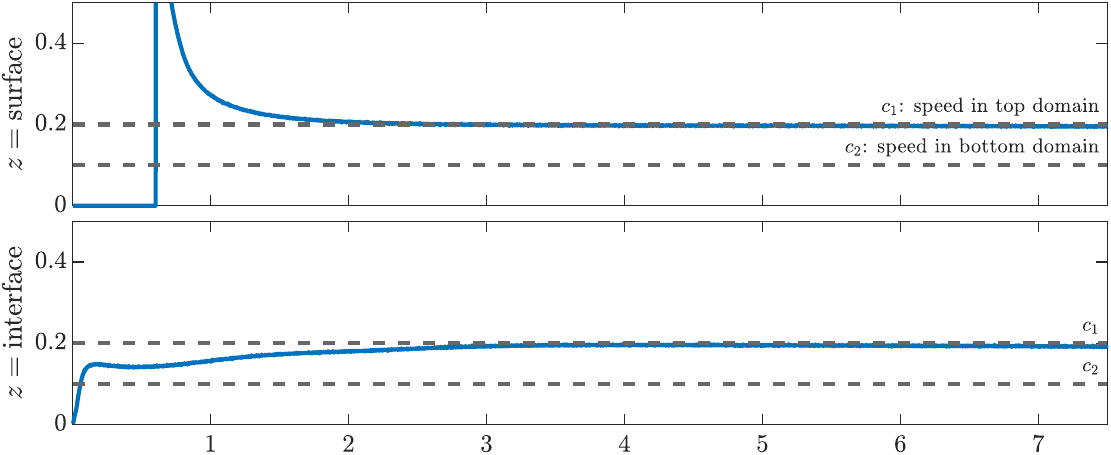}

\footnotesize
\centering

\caption{Speed of propagation along the interface and surface of Figure
	\ref{fig:gaussian_below_c1c2} over time.}
\label{fig:gaussian_below_c1c2_speed}
\end{figure}

\begin{figure}[htbp]
\includegraphics[width=\linewidth]{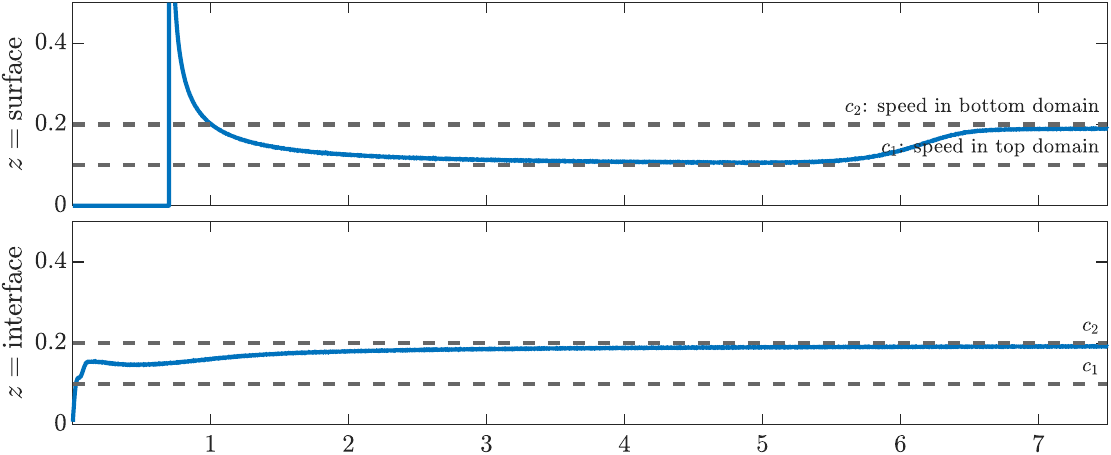}

\footnotesize
\centering

\caption{Speed of propagation along the interface and surface of Figure \ref{fig:gaussian_above_c1c2}
	over time. The outermost wave at the interface suddenly moves at speed $c_2$ around $t=6.5$ because
	that is when a faster moving wave is reflected from the surface of the domain passes the initial wave.}
\label{fig:gaussian_above_c1c2_speed}
\end{figure}

Figures \ref{fig:gaussian_below_c1c2_speed},
\ref{fig:gaussian_above_c1c2_speed}, and \ref{fig:gaussian_below_c2c1_speed}
plot the speed of the wave's propagation along the interface and the surface of the
simulations from figures \ref{fig:gaussian_below_c1c2},
\ref{fig:gaussian_above_c1c2}, and \ref{fig:gaussian_below_c2c1} respectively.
Recall the linear Love wave
existence condition \eqref{eqn:condition} which requires these wave speeds be
between $c_1$ and $c_2$. In every simulation we performed, the wave
propagating along the interface always tended to the maximum of
$\{c_1, c_2\}$. For the surface, the waves eventually tended to the maximum
of $\{c_1, c_2\}$. We see in Figure \ref{fig:gaussian_above_c1c2_speed} that
the wave speed along the surface is only \emph{initially} near the minimum of
$\{c_1, c_2\}$ when $c_1<c_2$ and the explosion originates from the upper
layer. The wave speed eventually tends to the larger wave speed because a
faster moving wave reflected from the surface becomes the outermost wave.
Some time must pass until the wave travels to the surface and to the interface,
and until then, the surface and interface velocities may be taken to be zero.
Changing of initial $z$-position and the spike width for the initial explosion
further from the interface results in corresponding different time of
surface/interface wave emergence. Because our interest lies mostly in studying
the wave's interactions with the interface, we save computational resources by
making the explosion close to the interface.

In Section \ref{sec:linearlove}, we derived a family of solutions to the linear
Love wave equation by assuming the phase of each individual mode is constant
with respect to $z$. This assumption does not agree with the results derived
from figures \ref{fig:gaussian_below_c1c2_speed} and
\ref{fig:gaussian_above_c1c2_speed}; however, it becomes more reasonable as the
simulations continue. Indeed, the set of solutions derived in Section
\ref{sec:linearlove} does not correspond to the initial conditions we
simulated.

%
%

To compute the wave speed along the surface (and the interface after making
minor changes to the following), we first find the largest value of $x_s$ such
that $\abs{u(x_s,L,t_s)}$ is greater than some numerical epsilon (in our case,
$10^{-3}$) for each nonuniform time step of the simulation $t_s$. The
velocity is then approximated with first-order forward differences
\[
	v_s \approx \frac{x_s - x_{s-1}}{t_s - t_{s-1}}.
\]
Rather than use the data $U_{\cdot,0}(t_s)\approx u(x,L,t_s)$ directly, we
build a spline from $U_{\cdot,0}(t_s)$ and evaluate the spline on $500$ times
more points. Using a spline is important because $x_{s+1}$ might
not change over the time $[t_s,t_{s+1}]$ if the mesh is not sufficiently
fine or the time step is too small. In each of our simulations, we already take
the step size $h$ as small as we can before our server runs out of RAM
($128$GB) and resorts to swap instead. As such, the spline roughly increases the
resolution of the data specifically along the surface.

We must ensure the adaptive time step $\Delta t_s = t_{s+1} - t_s$ is not too
small when computing $v_s$, because subtraction, and hence finite differences,
are not well conditioned. So, before plotting the wave speeds, we cull several
time steps so that the resulting $\Delta t_s$ is greater than some numerical
epsilon (in our case, $\Delta t_s > 10^{-3}$). Culling is especially important
in subsections \ref{sec:numerical_hyperelastic_love} and
\ref{sec:numerical_hyperviscoelastic_love} because the ODEs resulting from
discretizing the PDEs \eqref{eq:shear:genW:explicit} and
\eqref{eq:shear:genW:explicit:visc} are especially stiff during the initial
explosion. For our choice of parameters, such stiffness causes \texttt{ode23}
to take such small time steps that over $90\%$ of the time steps are culled.
We use this algorithm on $\abs{u(x_s,0,t_s)}$ to compute the wave speed along
the interface as well. This culling is exclusively for computing and plotting
the wave velocity.


\subsection{Nonlinear shear and Love-type waves}\label{sec:numerical_hyperelastic_love}

Now, we study the qualitative behaviour of the numerical solutions of equation
\eqref{Y1M6} (resulting from the cubic Yeoh strain energy density function
\eqref{eq:shear:genW:explicit}) applied to Love waves propagating through
hyperelastic materials,
\begin{equation}
	\label{eqn:eps}
	\frac{\rho_0}{2} v_{tt} = \div\left( [\mu(z) + M\norm{\grad v}^2 + A\norm{\grad v}^4]\grad v \right).
\end{equation}
In each experiment, we take $A=0$, ignoring the quintic terms. Again, we model
an isotropic layer overlying an isotropic elastic half-space by taking $\mu(z)$
to be a step function.

\begin{figure}[htbp]
\includegraphics[width=\linewidth]{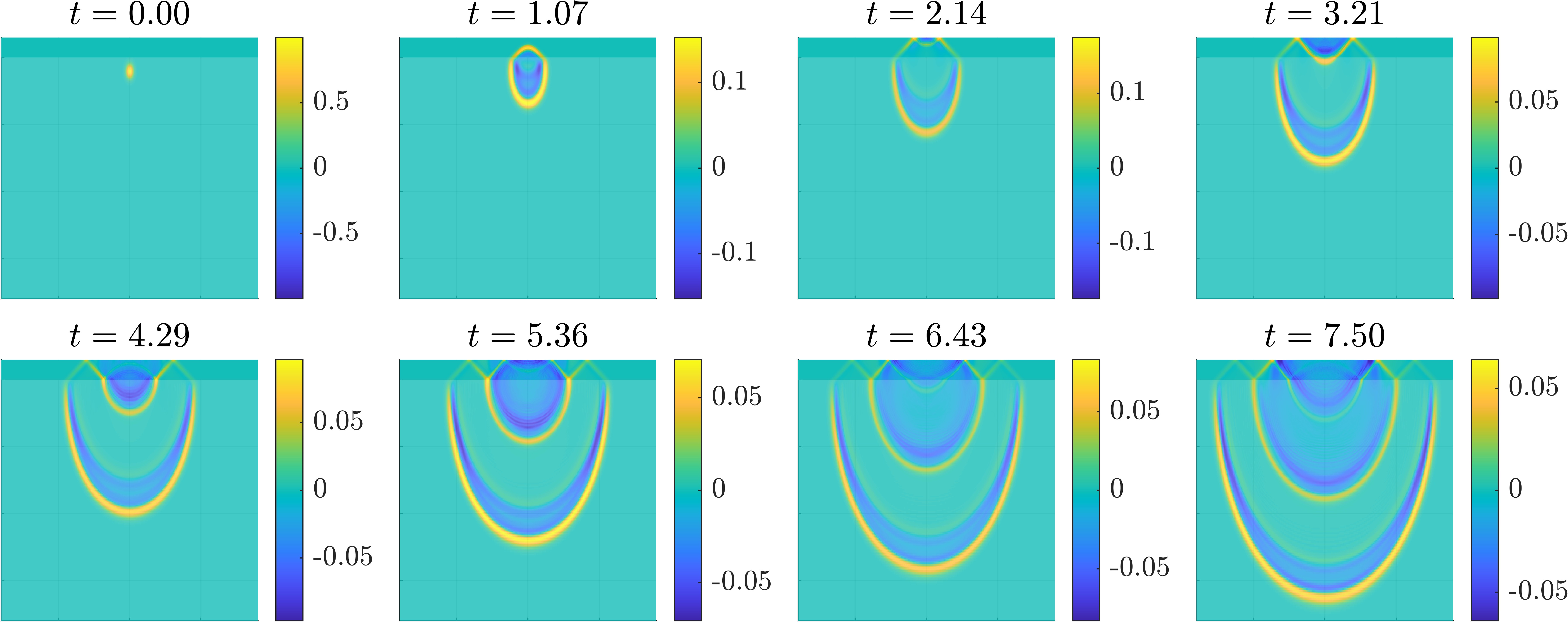}
\footnotesize
\centering

\caption{Plots of a numerical solution of \eqref{eqn:eps} with $c_1<c_2$ and $z_0$ in the lower half-space.
The parameters are $c_1 = 0.1$, $c_2 = 0.2$, $z_0=-0.1$, $r=0.05$, $A=1$, $L=0.15$, $M=\mathbf{5\cdot10^{-5}}$, $W=3.6$, $H=1.8$, and $h = \mathbf{0.0068}$.}
\label{fig:gaussian_below_c1c2_eps1}
\end{figure}

\begin{figure}[htbp]
\includegraphics[width=\linewidth]{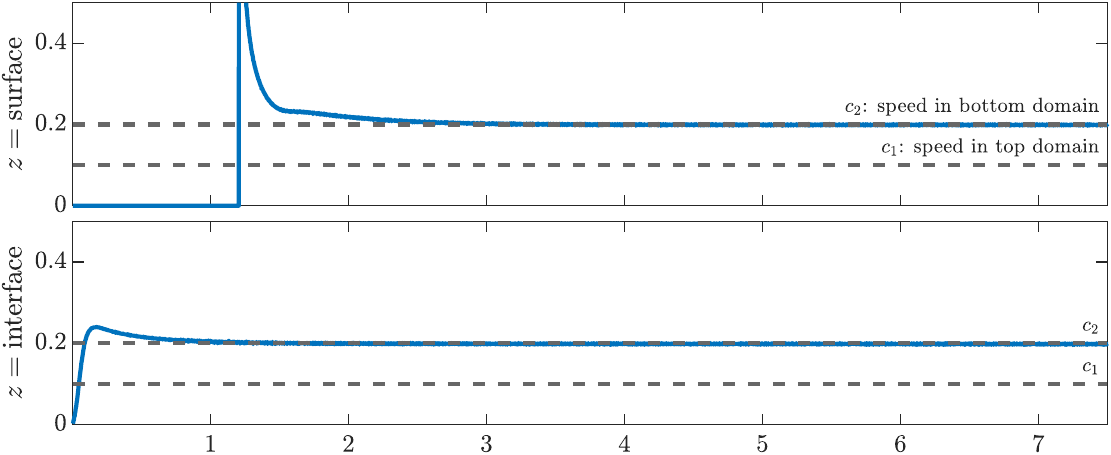}

\footnotesize
\centering

\caption{Speed of propagation along the interface and surface of Figure \ref{fig:gaussian_below_c1c2_eps1} over time.}
\label{fig:gaussian_below_c1c2_eps1_speed}
\end{figure}

\begin{figure}[htbp]
\includegraphics[width=\linewidth]{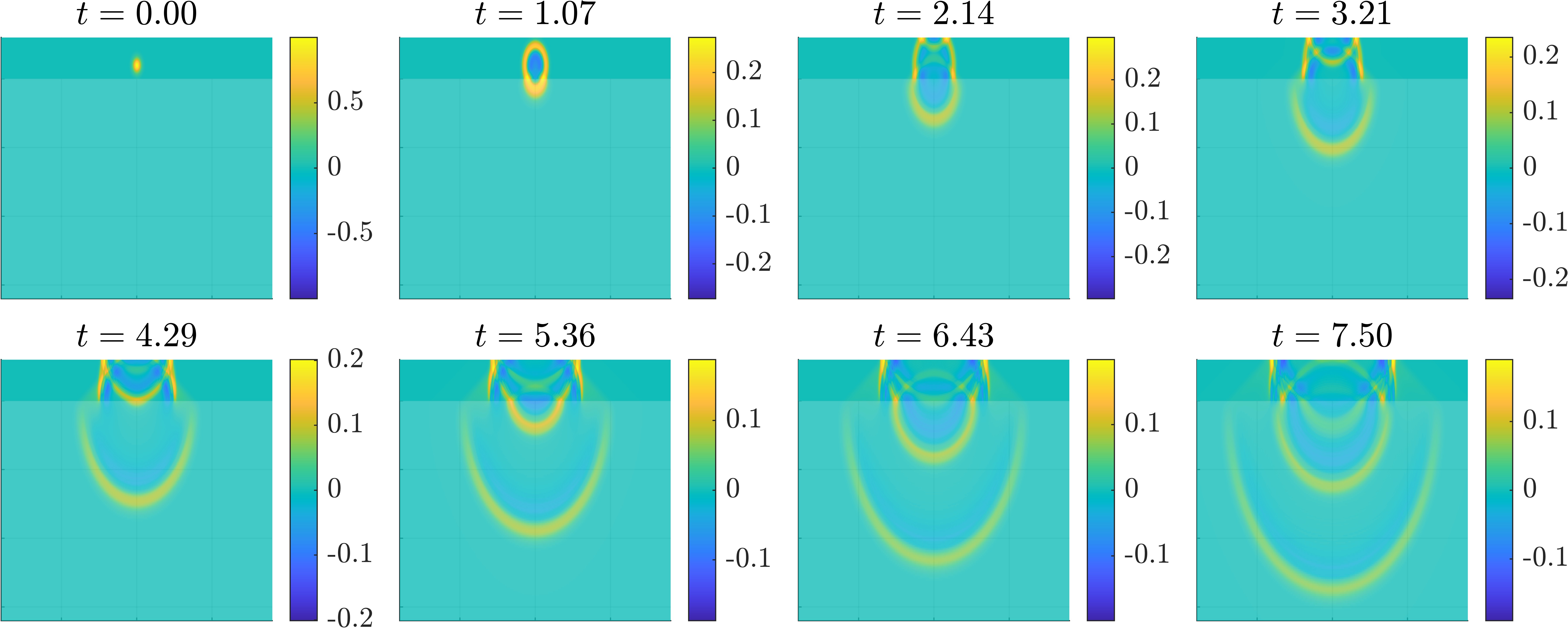}
\footnotesize
\centering

\caption{Plots of a numerical solution of \eqref{eqn:eps} with $c_1<c_2$ and $z_0$ in the upper layer.
The parameters are $c_1 = 0.1$, $c_2 = 0.2$, $z_0=\mathbf{0.1}$, $r=0.05$,
$A=1$, $L=\mathbf{0.3}$, $M=5\cdot10^{-5}$, $W=3.6$, $H=\mathbf{1.6}$, and $h = 0.0068$.}
\label{fig:gaussian_above_c1c2_eps1}
\end{figure}

\begin{figure}[htbp]
\includegraphics[width=\linewidth]{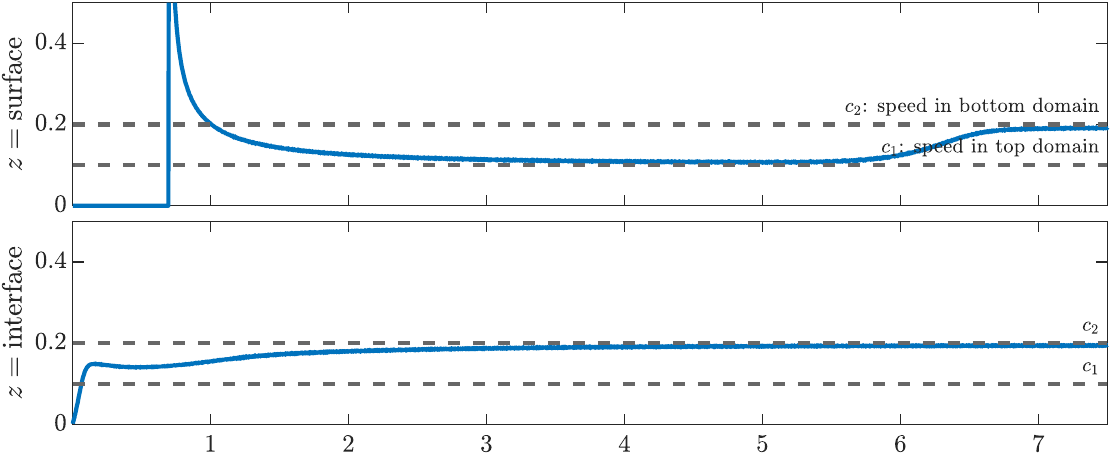}

\footnotesize
\centering

\caption{Speed of propagation along the interface and surface of Figure \ref{fig:gaussian_above_c1c2_eps1}.}
\label{fig:gaussian_above_c1c2_eps1_speed}
\end{figure}

\begin{figure}[htbp]
\includegraphics[width=\linewidth]{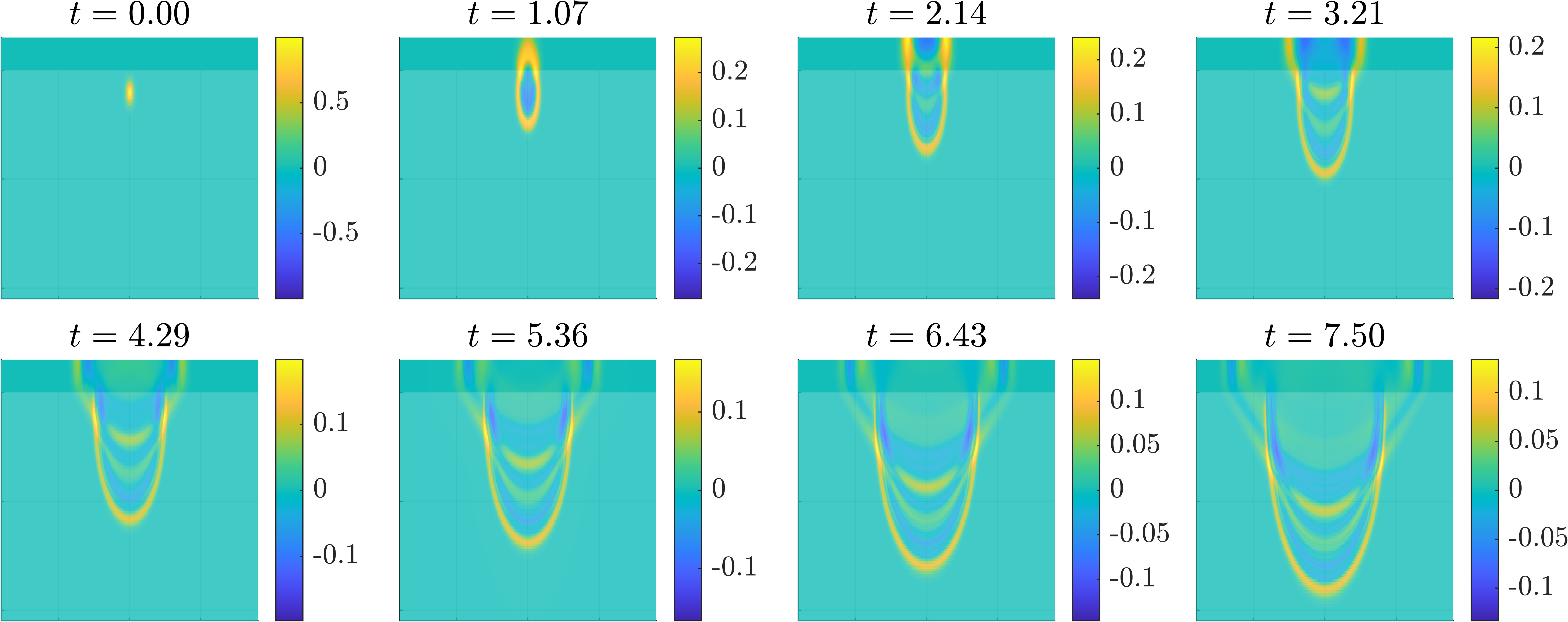}
\footnotesize
\centering

\caption{Plots of a numerical solution of \eqref{eqn:eps} with $c_1>c_2$ and $z_0$ in the lower half-space.
The parameters are $c_1 = \mathbf{0.2}$, $c_2 = \mathbf{0.1}$, $z_0=\mathbf{-0.1}$, $r=0.05$, $A=1$,
$L=\mathbf{0.15}$, $M=5\cdot10^{-5}$, $W=3.6$, $H=\mathbf{1.05}$, and $h = 0.0068$.}
\label{fig:gaussian_below_c2c1_eps1}
\end{figure}

\begin{figure}[htbp]
\includegraphics[width=\linewidth]{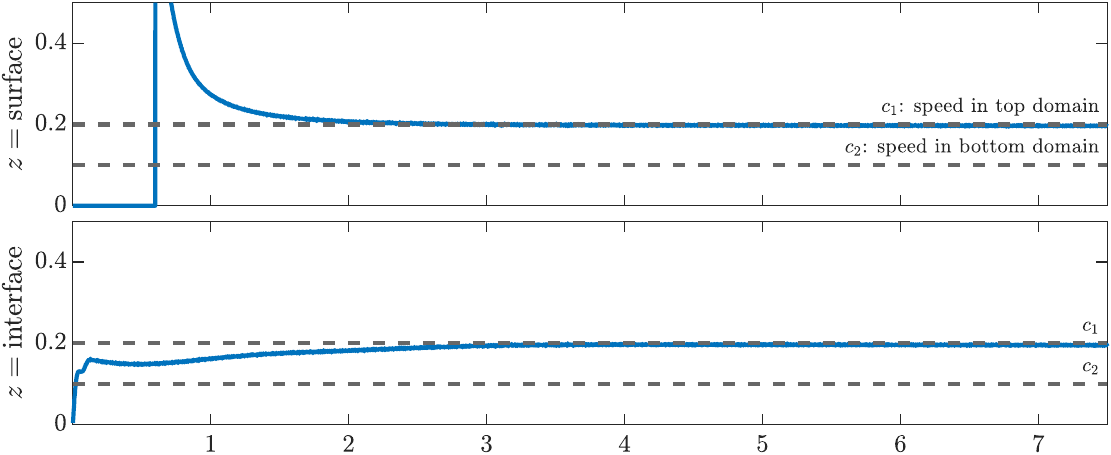}

\footnotesize
\centering

\caption{Speed of propagation along the interface and surface of Figure \ref{fig:gaussian_below_c2c1_eps1} over time.}
\label{fig:gaussian_below_c2c1_eps1_speed}
\end{figure}

Figures \ref{fig:gaussian_below_c1c2_eps1}, \ref{fig:gaussian_above_c1c2_eps1},
and \ref{fig:gaussian_below_c2c1_eps1} show the similar plots to those of the
previous subsection, but for the numerical solution of \eqref{eqn:eps} with
$M=5\cdot10^{-5}$. The solution's qualitative behaviour is very similar to
that of the linear case, even though $M$ was taken as large as possible within the
constraints of system memory. Increasing $M$ appears to result in a higher
frequency of small oscillations in the plots of $v(x,z,t)$. It is easier to see
these small oscillations after the explosion splits in all directions,
greatly decreasing the amplitude.
In the plots of the wave speeds (Figures
\ref{fig:gaussian_below_c1c2_eps1_speed},
\ref{fig:gaussian_above_c1c2_eps1_speed}, and
\ref{fig:gaussian_below_c2c1_eps1_speed}) we observe similar qualitative behaviour
as in the linear case shown above. In particular, there exists an initial period
of time where the velocity at the surface is zero;  the wave speed along the
surface is only initially the minimum of $\{c_1, c_2\}$ when $c_1<c_2$
and the explosion originates from the upper layer; the velocities at the
surface and interface eventually agree by tending to the  value of the maximum of $\{c_1, c_2\}$.

\subsection{Hyper-viscoelastic shear and Love-type waves}
\label{sec:numerical_hyperviscoelastic_love}
We now study the qualitative behaviour of the numerical solutions of equation
\eqref{eq:shear:genW:explicit:visc} (resulting from the cubic Yeoh strain energy
density function \eqref{eq:shear:genW:explicit} and the viscosity potential
\eqref{eq:Wv:our}) applied to Love waves propagating through hyper-viscoelastic
materials. In each experiment, we take $A=0$ and ignore any quintic effects in
the dispersion term $Q[v]$. The resulting model is
\begin{align} 
\rho v_{tt} =
	& 2\mu(v_{xx} + v_{zz}) \nonumber \\
	& + 2M\left(\left(v_x^2 + v_z^2\right)(v_{xx} + v_{zz})
		+ 2\left(v_x^2v_{xx} + v_z^2v_{zz} + 2v_xv_zv_{xz} \right)\right) \label{eqn:hypervisco} \\
	& + \eta \left(\left(v_x^2 + v_z^2\right)\left(v_{xxt} + v_{zzt}\right) \nonumber
		+ 2v_{xt}\left(v_x v_{xx} + v_z v_{xz}\right)
		+ 2v_{zt}\left(v_x v_{xz} + v_z v_{zz}\right)\right).
\end{align}

\begin{figure}[htbp]
\includegraphics[width=\linewidth]{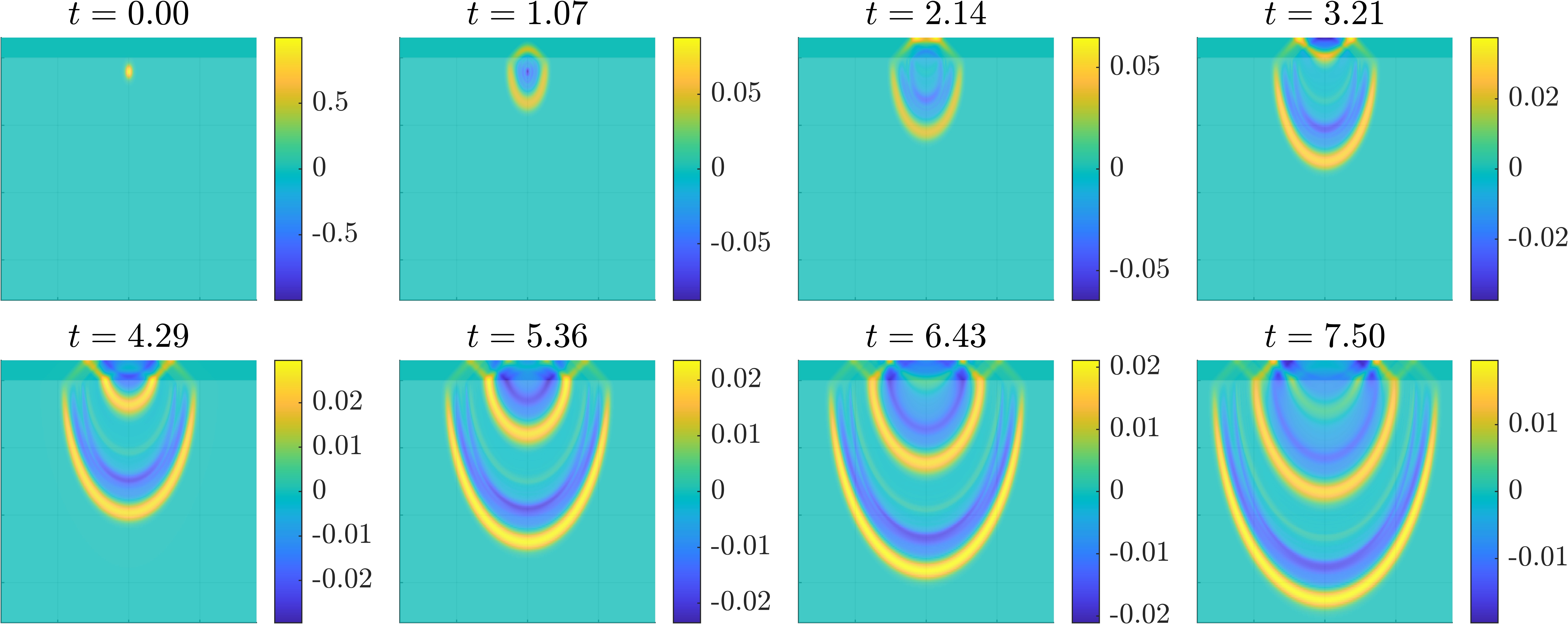}
\footnotesize
\centering

\caption{Plots of a numerical solution of \eqref{eqn:hypervisco} with $z_0$ in the lower half-space.
The parameters are $c_1 = 0.1$, $c_2 = 0.2$, $z_0=-0.1$, $r=0.05$, $A=1$,
$L=0.15$, $M=5\cdot10^{-5}$, $\eta=\mathbf{3.5\cdot10^{-4}}$, $W=3.6$, $H=1.8$,
and $h = 0.0068$.}
\label{fig:gaussian_below_c1c2_eps1_eta1}
\end{figure}

\begin{figure}[htbp]
\includegraphics[width=\linewidth]{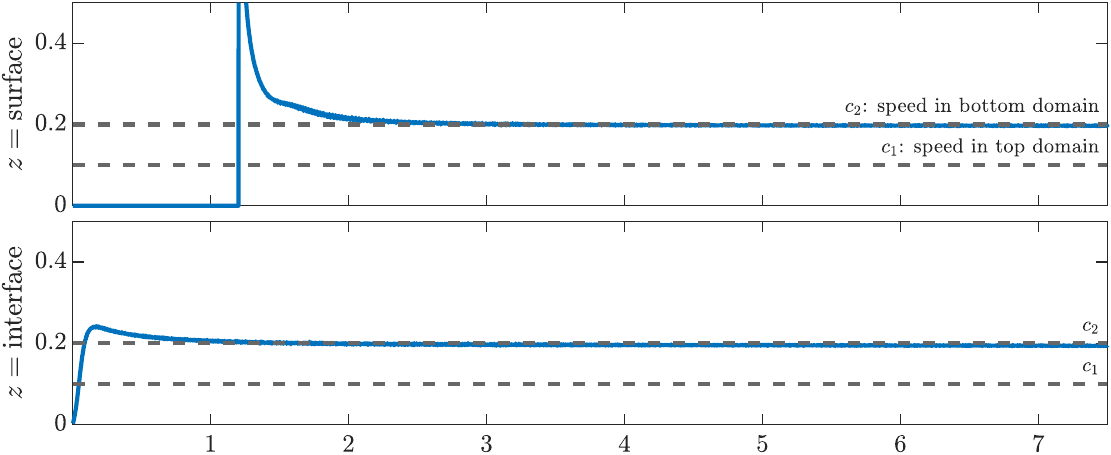}
\footnotesize
\centering

\caption{Speed of propagation along the interface and surface of Figure \ref{fig:gaussian_below_c1c2_eps1_eta1}.}
\label{fig:gaussian_below_c1c2_eps1_eta1_speed}
\end{figure}

\begin{figure}[htbp]
\includegraphics[width=\linewidth]{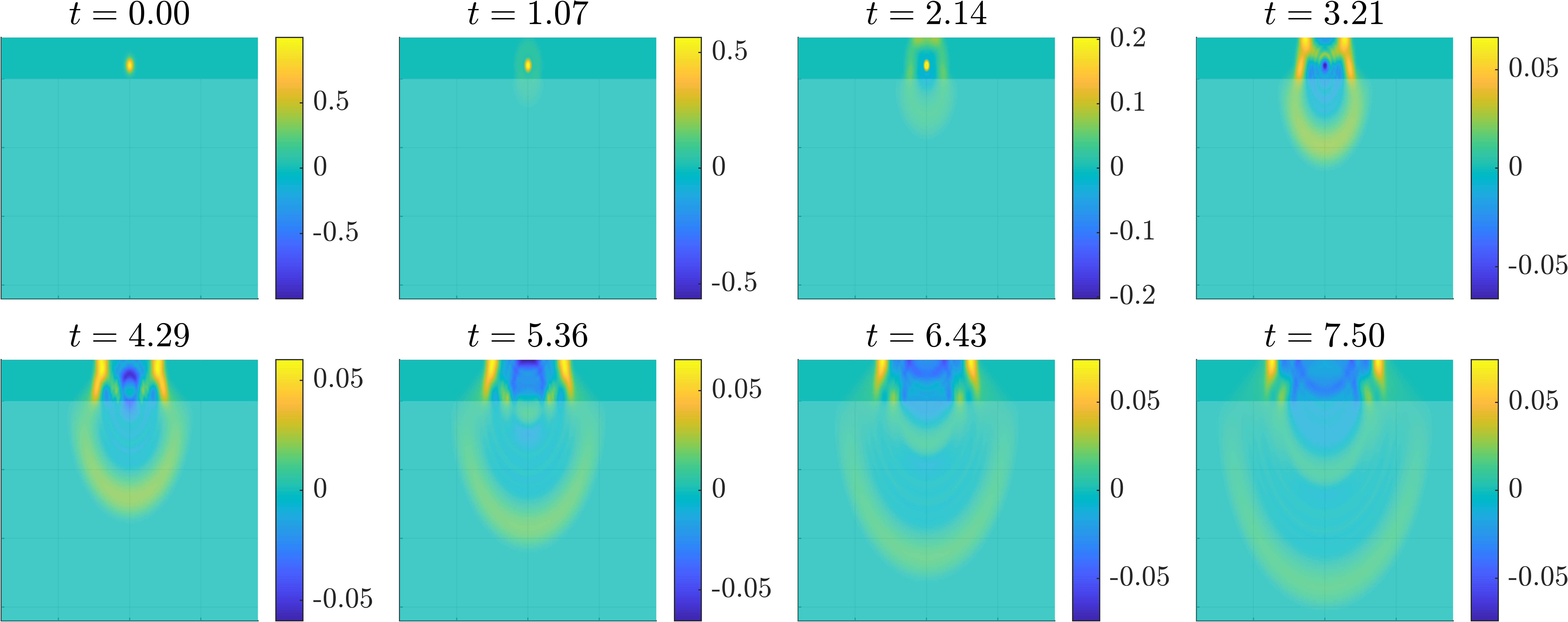}
\footnotesize
\centering
\caption{Plots of a numerical solution of \eqref{eqn:hypervisco} after a Gaussian
explosion centered in the upper layer. The parameters are $c_1 = 0.1$, $c_2 = 0.2$,
$z_0=\mathbf{0.1}$, $r=0.05$, $A=1$, $L=\mathbf{0.3}$, $M=5\cdot10^{-5}$,
$\eta=3.5\cdot10^{-4}$, $W=3.6$, $H=\mathbf{1.05}$, and $h = 0.0068$.}
\label{fig:gaussian_above_c1c2_eps1_eta1}
\end{figure}

\begin{figure}[htbp]
\includegraphics[width=\linewidth]{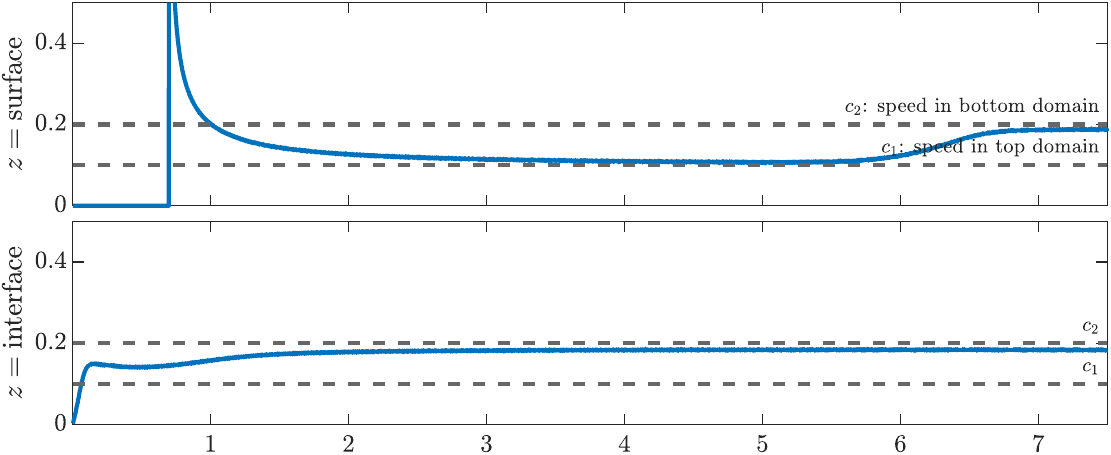}

\footnotesize
\centering

\caption{Speed of propagation along the interface and surface of Figure \ref{fig:gaussian_above_c1c2_eps1_eta1}.}
\label{fig:gaussian_above_c1c2_eps1_eta1_speed}
\end{figure}

\begin{figure}[htbp]
\includegraphics[width=\linewidth]{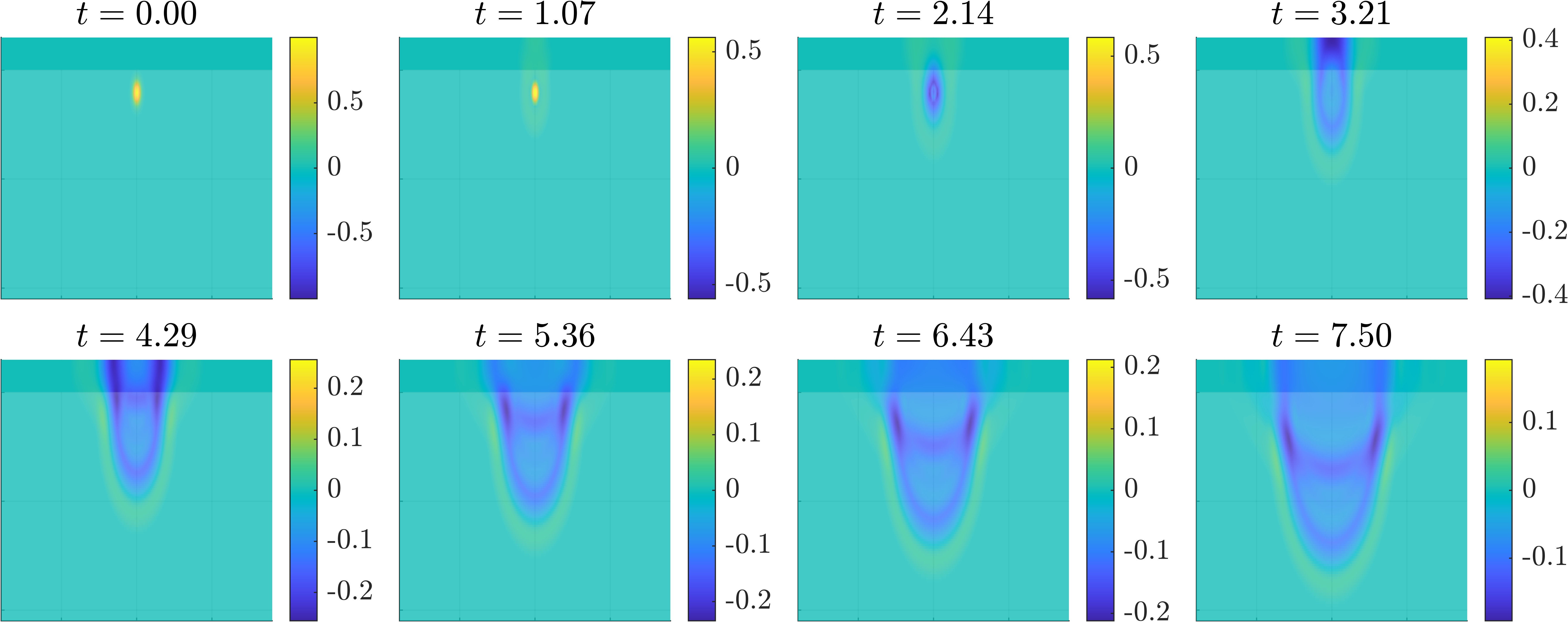}
\footnotesize
\centering

\caption{Plots of a numerical solution of \eqref{eqn:hypervisco} with $c_1>c_2$ after a Gaussian
explosion centered in the lower half-space. The parameters are $c_1 = \mathbf{0.2}$,
$c_2 = \mathbf{0.1}$, $z_0=\mathbf{-0.1}$, $r=0.05$, $A=1$, $L=\mathbf{0.15}$, $M=5\cdot10^{-5}$,
$\eta=3.5\cdot10^{-4}$, $W=3.6$, $H=1.05$, and $h = 0.0068$.}
\label{fig:gaussian_below_c2c1_eps1_eta1}
\end{figure}

\begin{figure}[htbp]
\includegraphics[width=\linewidth]{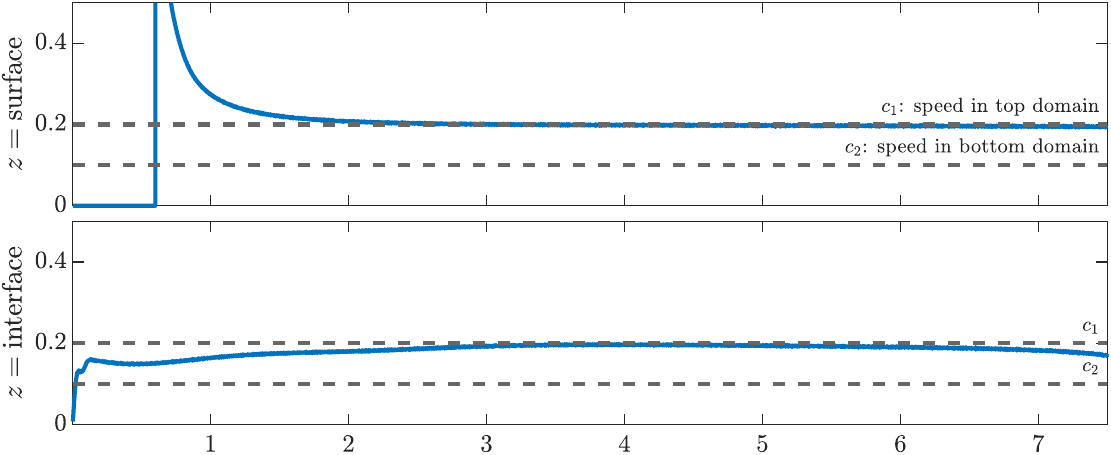}

\footnotesize
\centering

\caption{Speed of propagation along the interface and surface of Figure \ref{fig:gaussian_below_c2c1_eps1_eta1}.}
\label{fig:gaussian_below_c2c1_eps1_eta1_speed}
\end{figure}

Figures \ref{fig:gaussian_below_c1c2_eps1_eta1},
\ref{fig:gaussian_above_c1c2_eps1_eta1}, and
\ref{fig:gaussian_below_c2c1_eps1_eta1} show surface plots of a numerical
solution $v(x,z,t)$ of equation \eqref{eqn:hypervisco} with $M=5\cdot10^{-5}$
and $\eta=3.5\cdot10^{-4}$. The behaviour is again similar to that described in
the above two subsections; however, the waves disperse much more rapidly
throughout the domain. When $c_1<c_2$, the maximum value of $u(x,z,7.5)$ (the
point of largest displacement at the end of the simulation) is much less than
the equivalent simulation in the previous sections (likely due to the
dissipation of energy). When $c_2<c_1$ and the Gaussian explosion occurs below
the interface (Figure \ref{fig:gaussian_below_c2c1_eps1_eta1}), the
displacement $v(x,z,t)<0$ for more values of $x,z$ than any of the other
experiments. Our plots of the wave speeds
\ref{fig:gaussian_below_c1c2_eps1_eta1_speed},
\ref{fig:gaussian_above_c1c2_eps1_eta1_speed}, and
\ref{fig:gaussian_below_c2c1_eps1_eta1_speed} again shows similar behaviour to
the prior two subsections.

\begin{figure}[htbp]
	\includegraphics[width=\linewidth]{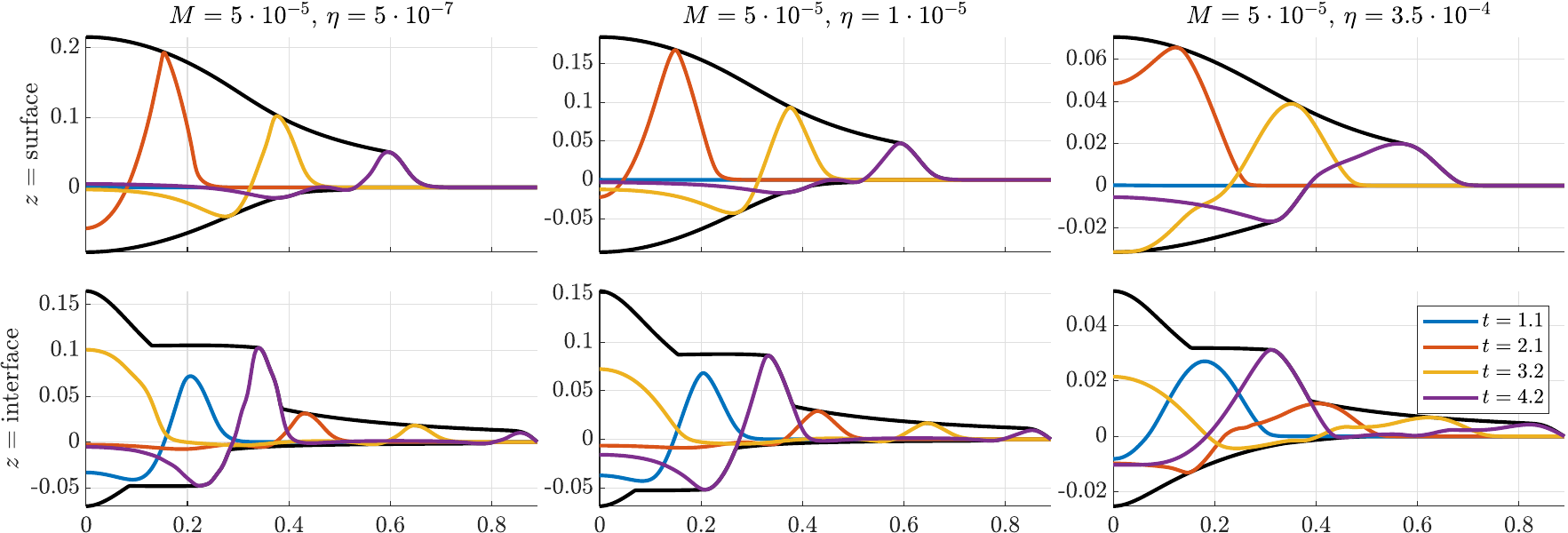}
\footnotesize
\centering

\caption{Plots of numerical solutions $v(x,z,t)$ of \eqref{eqn:hypervisco} over
$x$ (horizontal axis) for fixed $z$ along the interface and top of the domain
with $c_1<c_2$ after a Gaussian explosion in the lower half-space for various
values of $\eta$. The parameters are $c_1 = 0.1$, $c_2 = 0.2$, $z_0=-0.1$,
$r=1/16$, $A=0.3$, $L=0.15$, $W=2.1$, $H=2.45$, and $h = 0.068$.}
\label{fig:1Dvisco_waves}
\end{figure}

Figure \ref{fig:1Dvisco_waves} plots envelopes of the solution $v(x,0,t)$ and
$v(x,L,t)$ of equation \eqref{eqn:hypervisco} for four values of $t$ and four
values of $\eta$. We see the waves behave much more predictably as $\eta$
increases. For smaller $\eta$, the waves travelling at the surface appear to
skew to the right. In the sequel, we study the $1$D version of equation
\eqref{eqn:hypervisco}. In this $1$D setting, $v_{xt}(x,t)$ is unbounded when
$x$ is a local maximum of $v(x,t)$. This is not what we observe in the $2$D case.

\begin{figure}[htbp]
\includegraphics[width=\linewidth]{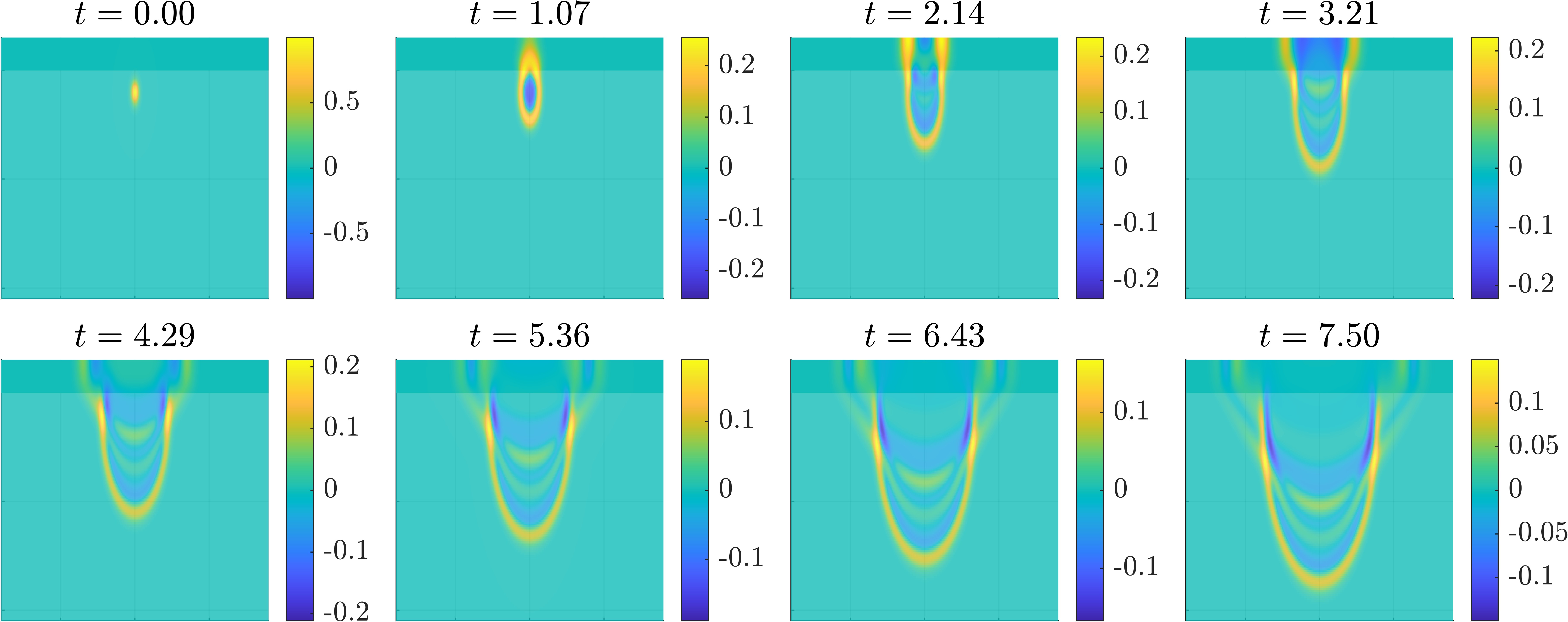}
\footnotesize
\centering

\caption{Plots of a numerical solution of \eqref{eqn:linearlove} with $c_1>c_2$
and $z_0$ in the lower half-space.  The parameters are $c_1 = \mathbf{0.2}$,
$c_2 = \mathbf{0.1}$, $z_0=\mathbf{-0.1}$, $r=0.05$, $A=1$, $L=\mathbf{0.15}$,
$W=3.6$, $H=\mathbf{1.05}$, and $h = 0.005$.}
\label{fig:gaussian_below_c2c1}
\end{figure}

\begin{figure}[htbp]
\includegraphics[width=\linewidth]{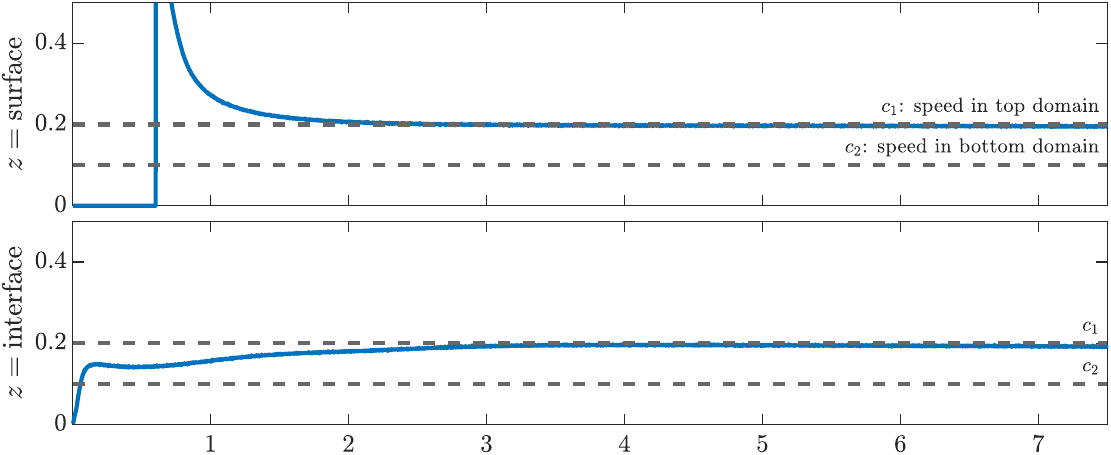}

\footnotesize
\centering

\caption{Speed of propagation along the interface and surface of Figure \ref{fig:gaussian_below_c2c1} over time.}
\label{fig:gaussian_below_c2c1_speed}
\end{figure}

\newpage
\section{Symmetry analysis of the one-dimensional model}\label{sec:1d}
We are interested in applying the Lie symmetry method to the $1$D version of
equation \eqref{eqn:hypervisco}, given by
\begin{equation}\label{eq:1d:visco}
	\rho u_{tt} = 2(\mu + 3M u_x^2) u_{xx} + \eta (u_{xt} u_x^2)_x.
\end{equation}
The Lie symmetries of a PDE, when they exist, are of interest because they
provide an algorithmic way of deriving special sets of exact solutions, namely, invariant solutions. They
are one of few methods for deriving solutions to nonlinear PDEs. Further, many
familiar ans\"{a}tze (such as the travelling wave ansatz or the self-similar ansatz)
are special cases of Lie symmetries.

\begin{figure}[htbp]
\includegraphics[width=\linewidth]{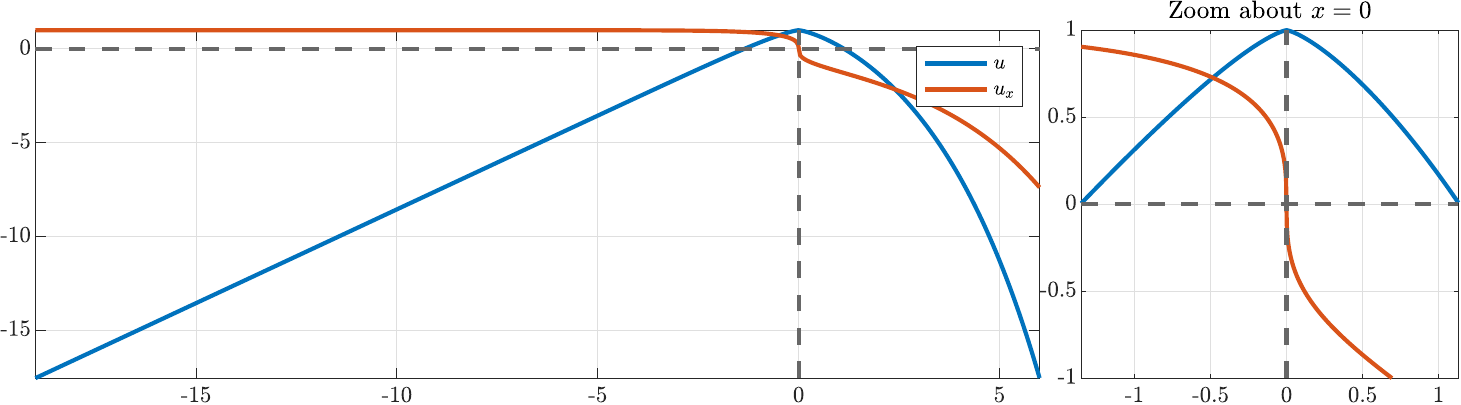}
\footnotesize
\centering

\caption{Solution \eqref{eqn:symmsoln} and its spatial derivative using
$c_1 = 1$, $c_2 = -1$, and $c_3 = -c_2$. When comparing this to other plots, note that for the travelling wave with $c=1$, $u_x = -u_t$.}
\label{fig:symmetry_eta}
\end{figure}

\begin{figure}[htbp]
\includegraphics[width=\linewidth]{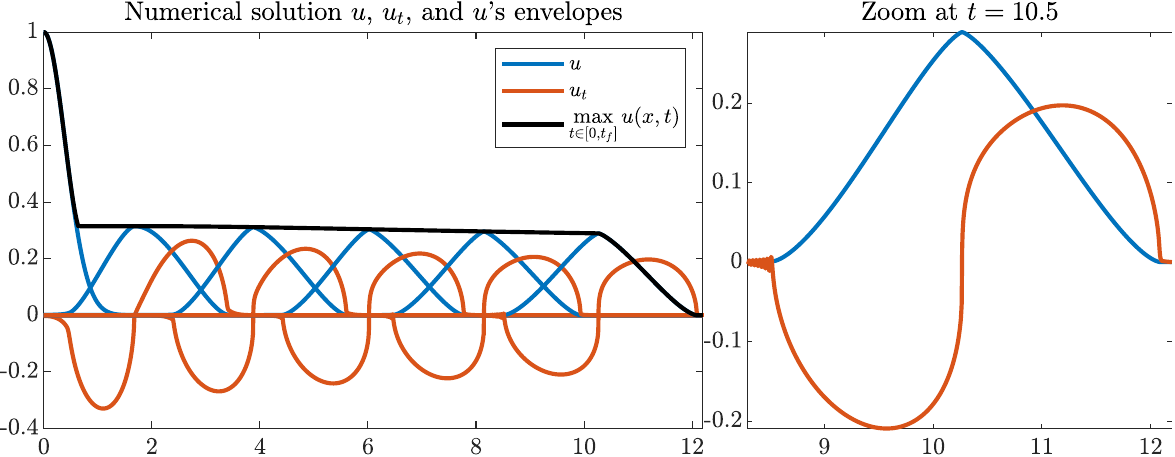}
\footnotesize
\centering

\caption{Left: the dynamics of the numerical solution $u(x,t)$ and $u_t(x,t)$ of equation \eqref{eqn:1d_eta}
	for various $t$. Initially there is a Gaussian explosion at $x=0$, after
	which the system splits into symmetric left- and right-travelling waves; only the
	right half is shown. The $u$ envelope is plotted in black.  Right: A
zoomed-in plot of the final values of $u$ and $u_t$.}
\label{fig:1D_eta_gaussian}
\end{figure}

Equation \eqref{eq:1d:visco} can be nondimensionalized by dividing both sides
by $\rho\neq0$ and setting
\begin{gather*}
	t = \frac{\eta}{6M}T, \quad x = \frac{\eta}{M}\sqrt{\frac{\mu}{18}}X, \quad u(x,t) = \frac{\eta\mu}{\sqrt{54M^3}}U(X,T).
\end{gather*}
The resulting equation becomes
\begin{equation}
	\label{eqn:1d_eta}
	U_{TT} = (1 + U_X^2)U_{XX} + (U_{XT}U_X^2)_X = U_{XX} + U_X^2U_{XX} + (U_X^2U_{XX})_T.
\end{equation}
It is straightforward to use a computer algebra software package to compute
each point symmetry of \eqref{eqn:1d_eta}. The four symmetry generators of
\eqref{eqn:1d_eta} as computed via the implementation of Lie's algorithm in the
\verb|Maple| package \verb|GeM| \cite{cheviakov2007gem, cheviakov2010symbolic}
are
\[
	X_1 = \pdv{x}, \quad X_2 = \pdv{t},\quad  X_3 = \pdv{u},\quad  X_4 = t\pdv{u}.
\]
Because Lie's algorithm is complete, these four generators form a basis of the
Lie algebra of point symmetries of \eqref{eqn:1d_eta}. Several families of
solutions to equation \eqref{eqn:1d_eta} can be derived by studying the
invariant curves of a linear combination of these symmetry generators. The
first integrals of
\begin{equation}\label{eq:symm:combination}
	X = c\pdv{x} + \pdv{t} + (k_1 + tk_2)\pdv{u},
\end{equation}
for constants $c,k_1,k_2\in\mathbb{C}$ are $I_1 = x-ct$ and $I_2 = k_1t + k_2t^2/2 - u$,
so one family of solutions is of the form $u(x,t) = k_1t + k_2t^2/2 + F(x-ct)$,
where $F$ is any solution to the ODE
\begin{equation}
	\label{eqn:1D_travel}
	cF'^2F''' = F''(F'^2 - 2cF'F'' + 1 - c^2) - k_2.
\end{equation}
If we require $u(x,t)$ to be bounded over $t$, then $k_1 = k_2 = 0$ and
equation \eqref{eqn:1D_travel} simply describes the travelling wave solutions
of the PDE \eqref{eqn:1d_eta}. It is straightforward to derive an exact solution when $c=\pm1$. In this case,
the ODE becomes ${F'''}/{F''} + 2{F''}/{F'} = \pm 1$, which yields the family of exact solutions
\begin{equation}
	\label{eqn:symmsoln}
	F(x) = c_1 + \int_0^x \sqrt[3]{c_2e^{\sign(c)w} + c_3} \, \mathrm{d}w
\end{equation}
for some real constants $c_1,c_2,c_3$. A physically realizable wave-type
travelling wave solution must be bounded and possess a local extremum. No
choice of $c_1,c_2,c_3$ leads to a bounded solution over all of $\mathbb{R}$;
however, one endpoint ($\pm\infty$) can be bounded by taking $c_3=0$. A critical point
$x_c$ of $F$ only exists when $c_2$ and $c_3$ have different signs. Such $x_c$
is a local maximum or minimum because $F''(x_c) = \pm\infty$. Figure \ref{fig:symmetry_eta}
plots such a solution $F$. For comparison, Figure \ref{fig:1D_eta_gaussian}
uses the method of lines to plot a numerical solution of \eqref{eqn:1d_eta} for
various times after a Gaussian explosion. The numerical solution quickly
evolves to look like a bounded version of equation \eqref{eqn:symmsoln} (where
$u_{tx}$ is unbounded at the maximum of $u$).

\begin{figure}[htbp]
	\includegraphics[width=\textwidth]{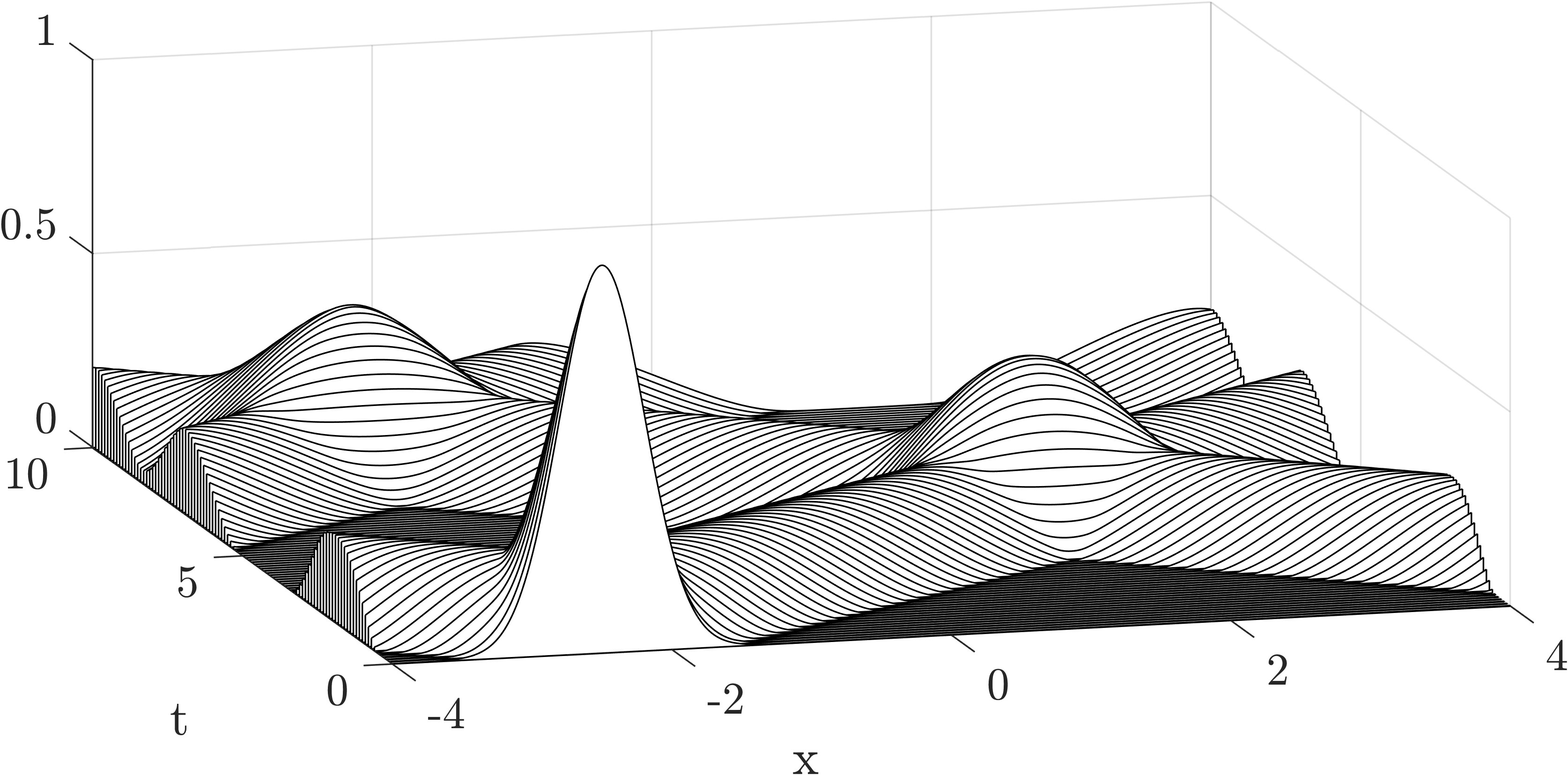}
\footnotesize
\centering

\caption{The travelling wave amplitude $u(x,t)$ in a domain with periodic boundary
	conditions for an initial Gaussian explosion $u(x,0) = e^{-\ln(100)(x^2+2.5)/0.85^2}$.}
\label{fig:collisions}

\end{figure}

The shape in the numerical solution in Figure \ref{fig:1D_eta_gaussian} is rather
stable. To demonstrate its stability, we re-solved the above problem
on a domain with periodic boundary conditions, took the time span long enough
for the wave to collide with itself twice, and plotted the results in Figure
\ref{fig:collisions}. The nonlinear and dispersive effects are much less apparent
over time as the wave amplitude decreases, effectively switching the system into the linear mode.

\begin{figure}[htbp]
\includegraphics[width=\linewidth]{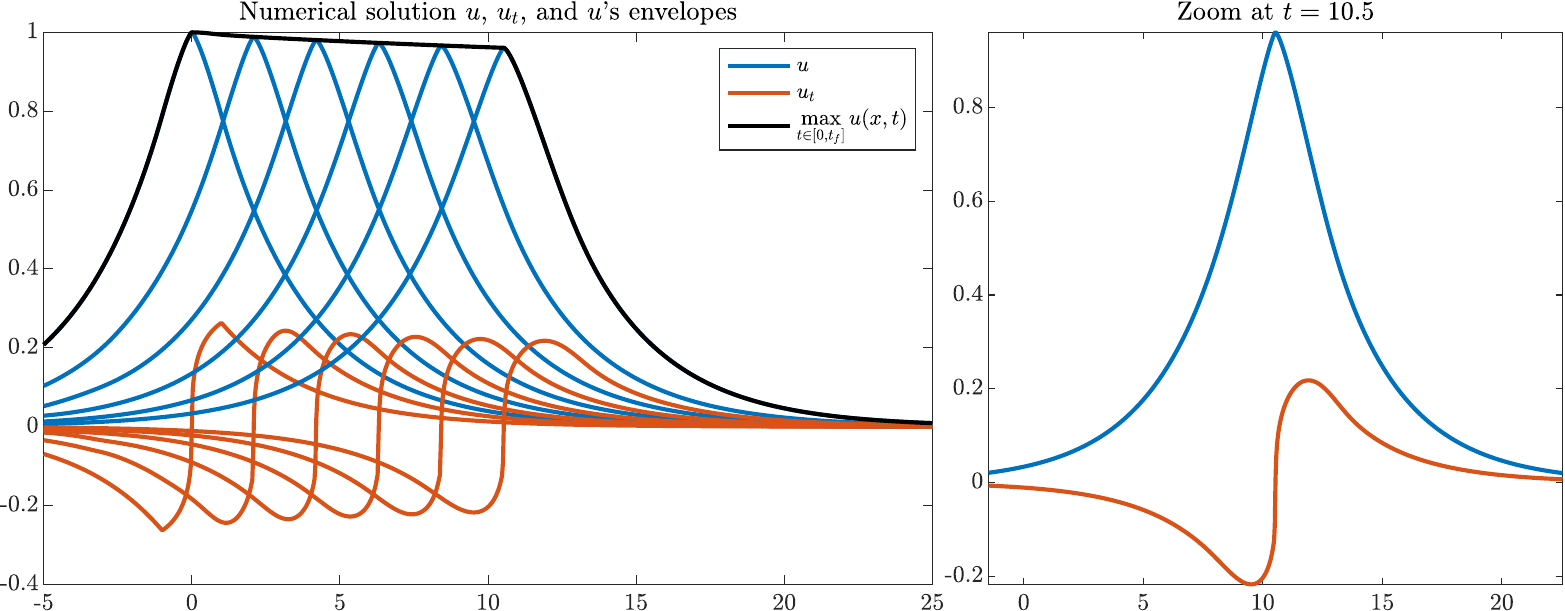}
\footnotesize
\centering

\caption{Left: plots of the solutions $u(x,t)$ and $u_t(x,t)$ of equation \eqref{eqn:1d_eta} for various $t$.
The initial condition is a piecewise continuously differentiable function whose pieces are of the form \eqref{eqn:symmsoln}.
Initially, $u_t$ is not differentiable at the breakpoints, but smooths out as the simulation progresses.}
\label{fig:1D_eta_symm}
\end{figure}

We note that it is possible to construct a bounded continuously differentiable piecewise
function as a combination of the symmetry invariant solutions \eqref{eqn:symmsoln} taking, for example,
\[
	(F'(x))^3 = a
	\begin{cases}
		e^{-x}-1      & \text{if } 0 \leq x \leq b, \\
		(1-e^b)e^{-x} & \text{if } b < x, \\
		-(F'(-x))^3       & \text{otherwise}.
	\end{cases}
\]
Though this formula does not yield a travelling wave solution itself, it produces a well-behaved wave when used as an initial
condition for the PDE \eqref{eqn:1d_eta} along with the time derivative condition
$u_t(x,t) = -cF'(x)$. The numerical solution is plotted in Figure
\ref{fig:1D_eta_symm} along with the wave's envelope.

\section{Discussion}\label{sec:discussion1}

Waves travelling on interfaces between materials with different mechanical properties, such as Love waves in geological sciences, are an important object of study due to their unique propagation properties. In this paper, hyperelasticity and hyper-viscoelasticity frameworks were used to derive equations for shear displacements in the $Y$-direction propagating in $(X,Z)$-plane, holding for an arbitrary hyperelastic potential (equation \eqref{eq:shear:genW:explicit}). With these frameworks, the linear wave equations for the Love wave model of interfacial waves was generalized to the nonlinear case, in particular, for the cubic Yeoh constitutive model \eqref{eq:n1} (equation \eqref{eq:y1M}).

In the context of fiber-reinforced anisotropic solids, it has been previously shown that one-dimensional nonlinear waves of the form \eqref{eqn:nablas} with purely quadratic nonlinearity ($A=0$) led to wave breaking \cite{cheviakov2016one}, which was eliminated when a viscosity term was taken into account. In this work, viscoelastic effects in one and two spatial dimensions were described through the pseudopotential \eqref{eq:Wv:our} and the corresponding the viscoelastic stress \eqref{eq:PK2:viscoel}.

Nonlinear Love-type waves propagating along the contact of two solid media (Figure \ref{fig:setup}) satisfying the Love wave existence condition $c_1 < \abs{v} < c_2$ \eqref{eq:c1c2cL} and the linear Hookean, nonlinear hyperelastic, and nonlinear hyper-viscoelastic equations were studied using the numerical method of lines in two spatial dimensions (Section \ref{sec:numerical}). Interfacial waves on the contact of the two media and surface waves at the free surface of the top domain were tracked and their speeds were determined; in agreement with the linear Love wave theory, it was found that the phase velocity of the nonlinear Love waves satisfied the condition \eqref{eq:c1c2cL}, with the phase velocity approaching the larger of the two material velocities $c_2$ (Figures \ref{fig:gaussian_below_c2c1}--\ref{fig:gaussian_below_c2c1_eps1_eta1_speed}).

For a one-dimensional reduction, the nonlinear wave equations \eqref{eq:1d:visco} in hyper-viscoelastic settings were studied using the point
symmetry analysis. The linear combination of symmetries \eqref{eq:symm:combination} yielded an invariant solution of a physical wave shape; however, that solution was unbounded. Numerical simulations with appropriate initial conditions demonstrated a qualitative agreement of numerical and exact solutions, in particular, the propagation of a wave with approximately constant speed and an amplitude decaying due to viscous effects.

Future work may include the study of anisotropic effects of Love-type wave
behaviour through the use of more general hyperelastic constitutive laws with
additional anisotropy terms (e.g., \cite{cheviakov2016one, cheviakov2021radial}
and references therein). It is also of interest to apply hyperelasticity theory
to derive and study nonlinear generalizations of equations for other
(compressional) surface waves, including Rayleigh waves that propagate along
the free surface between a vacuum overlaying a solid, Scholte waves that move
along fluid-solid interfaces, and Stoneley waves that propagate along the
interface between two elastic solids. The analysis of such waves is complicated
by the need to use compressible hyperelasticity, which employs a different set
of strain tensor invariants and results in significantly more complex equations
of motion.

\subsection*{Acknowledgements}

S.M. and S.A. thank the Department of Mathematics and Statistics, University of Saskatchewan for the support throughout the Master's thesis program. S.M. and A.C. are grateful to NSERC of Canada for research support through a CGS-M grant and a Discovery grant RGPIN-2024-04308.

\section*{Addendum}
The original version of this paper contains an error in Subsection
\ref{sec:nonlWderivation}, where we substituted the result
$I_1=I_2=v_X^2+v_Z^2+3$ into the strain energy density function before
expanding the formula for the Piola-Kirchoff tensor $\tens{P}$ in
\eqref{eq:PCformulas}. In effect, we implicitly assumed that $W^H_2=0$,
neglecting the compatibility condition \eqref{eq:compatibility} necessary to
ensure the hydrostatic pressure $p$ is well defined when $W^H_2\neq0$. This led
to some minor errors in other examples. These issues are remedied by explicitly
stating our focus on generalized neo-Hookean materials. These are materials
where the strain energy density function $W^H$ does not depend on the second
invariant: $W^H_2=0$. This error led to some other minor errors throughout
which are now corrected. Because of our restriction to generalized neo-Hookean
materials, any discussion around the Murnaghan model was replaced with the
cubic Yeoh model. There was one other minor error in the caption of Figure
\ref{fig:collisions}, which is now corrected. The remainder of the paper stands
without issue.

Since publishing, we discovered numerically breaking wave solutions, which we
include as an appendix. The numbering for each section, figure, and equation
was adjusted to include this work as a chapter in this thesis.


\appendix

\section{A comment on radially symmetric wave solutions that break numerically}
\begin{figure}[thbp]
\includegraphics[width=\linewidth]{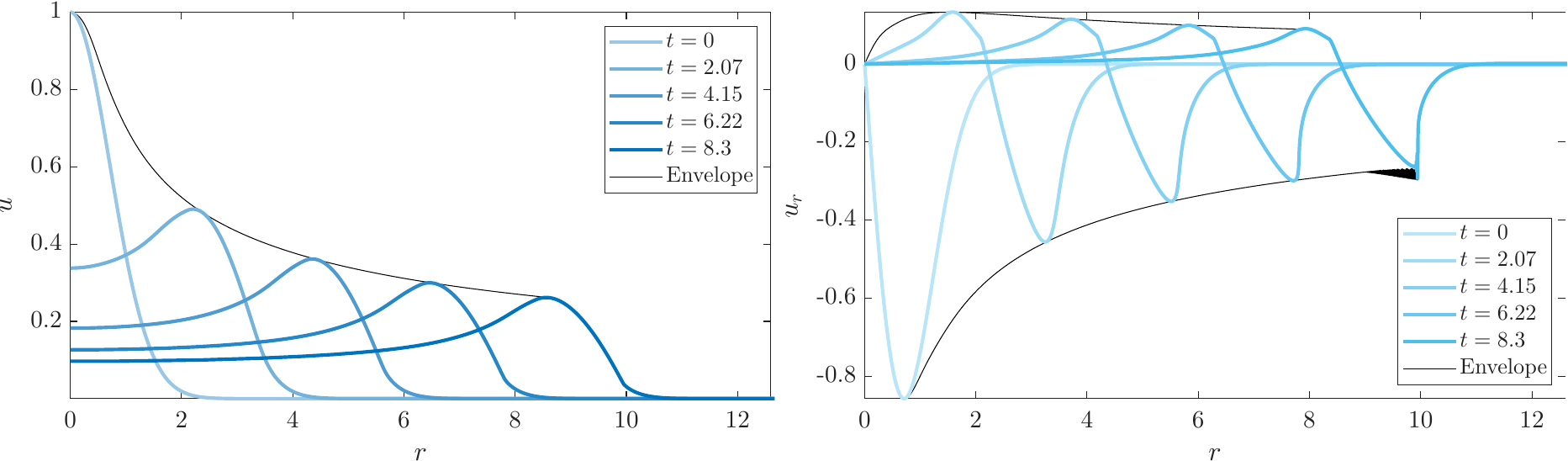}
\footnotesize
\centering
\caption{Profiles of a numerical solution of \eqref{eq:radially_symmetric}
(left) and its spatial derivative $u_r$ (right) at various times as computed
via the method of lines for $u(r,0)=e^{-r^2}$, $u_t(r,0)=-\dv{r} u(r,0)$,
$W'(I_1)=a+\epsilon I_1$, and $\epsilon=1$. The numerical domain is
$[0,12.6]$ discretized by a uniform mesh of step size $h=0.00105$.}
\label{fig:breaking_wave}
\end{figure}

At the time of publishing, the authors had not observed numerically breaking
wave solutions of the PDE
\begin{equation}
	\label{eq:appendix_eq}
	u_{tt} = \nabla\cdot(W'(u_X^2+u_Z^2)\nabla u).
\end{equation}
Equation \eqref{eq:appendix_eq} is \eqref{eq:shear:genW:y} with each constant
absorbed into $W$. Since then, we discovered that it is possible for such
solutions to break numerically. To study these solutions at a high resolution,
we assert that the initial conditions and domain are radially symmetric. This
assertion results in the $(1+1)$-dimensional PDE
\begin{equation}
	\label{eq:radially_symmetric}
	ru_{tt} = (rW'(u_r^2)u_r)_r,
\end{equation}
where $r$ is the radius from the point of symmetry. Taking $W'(I_1)=a+\epsilon
I_1$ yields the equation
\begin{equation}
	\label{eq:twoD}
	ru_{tt} = (r(u_r+\epsilon u_r^3/3))_r, \quad u_r(0,t) = 0, \quad u(\infty,t) = 0
\end{equation}
Figure \ref{fig:breaking_wave} plots profiles of \eqref{eq:twoD} at various
times as computed numerically via the method of lines with initial conditions
$u(r,0)=e^{-r^2}$, $u_t(r,0)=-\dv{r} u(r,0)$, and $\epsilon=1$. For this mesh,
spurious oscillations form in $u_r$ as early as $t=6.8$, and begin to alter the
envelope of $u_r$ around $t=7.5$. These oscillations are typical for
numerically breaking waves. The solution $u$ itself remains continuous, but a
sudden change in its slope is evident around $t=8.3$ and $r=10$. After the wave
passes, the material remains displaced by at least $0.1$ for all simulation
times and appears to gradually return to its initial configuration.

To compare against another method of reducing the equation's dimension, assume
$u_Z=0$ in \eqref{eq:appendix_eq} to see
\begin{equation}
	\label{eq:zero_Z}
	u_{tt} = (W'(u_X^2)u_X)_X.
\end{equation}
Symbolically, this equation is similar to \eqref{eq:radially_symmetric}, but
solutions of these two equations behave very differently. For example, the
boundary conditions are different and \eqref{eq:zero_Z} is missing a factor of
the independent variable $X$. Taking $W'(I_1)=a+\epsilon I_1$ yields the
equation
\begin{equation}
	\label{eq:oneD}
	u_{tt} = (u_X+\epsilon u_X^3/3)_X.
\end{equation}
The spatial derivative $u_X$ forms a jump discontinuity around $t=1.507$ for
the same initial conditions and $\epsilon=1$. It appears that solutions to
\eqref{eq:oneD} typically break sooner than those of \eqref{eq:twoD}.

{\footnotesize
\bibliography{shear_love_waves_12}
\bibliographystyle{ieeetr}
}

\end{document}